\documentclass[draft]{agujournal2019}
\usepackage{url} 
\usepackage{lineno}
\usepackage{soul}
\usepackage{amsmath}
\usepackage{amssymb}
\draftfalse

\journalname{}

\begin{document}

\title{Stress state, subsidence, and faulting in the Wilmington Oil Field, California: a multiphase flow--geomechanics modeling assessment (1936--2020)}
\authors{Lluís Saló-Salgado\affil{1,2,3}, Josimar A. Silva\affil{1}\thanks{Current address: ExxonMobil Technology and Engineering Company, Spring, TX, 77389}, Andreas Plesch\affil{1}, Franklin D. Wolfe\affil{1}, John H. Shaw\affil{1}, Ruben Juanes\affil{2,4}}

\affiliation{1}{Dept. of Earth and Planetary Sciences, Harvard University, Cambridge, MA 02138}
\affiliation{2}{Dept. of Civil and Environmental Engineering, Massachusetts Institute of Technology, Cambridge, MA 02139}
\affiliation{3}{Dept. of Earth, Environmental, and Planetary Sciences, The University of Tennessee, Knoxville, TN 37996}
\affiliation{4}{Dept. of Earth, Atmospheric and Planetary Sciences, Massachusetts Institute of Technology, Cambridge, MA 02139}

\correspondingauthor{John H. Shaw}{shaw@eps.harvard.edu}
\correspondingauthor{Ruben Juanes}{juanes@mit.edu}

\begin{keypoints}
\item We combine multimodal field data and coupled flow--geomechanics modeling to assess the impact of reservoir operations on fault stability
\item Our results suggest that a critically-stressed, reverse faulting stress regime in the field is unlikely and needs to be reassessed
\item Intra-reservoir faults and, especially, shallow sub-horizontal surfaces were destabilized, consistent with documented well damage
\end{keypoints}

\begin{abstract}
     Nearly a century of oil production in the Wilmington Oil Field, Los Angeles Basin, California, has modified the stress state, caused nearly 9 m of ground surface subsidence, and been associated with earthquakes that sheared wells. This offers a unique opportunity to elucidate the processes that govern these phenomena: Since the 1930s, approximately 2.5 billion barrels of oil have been produced, accompanied by water injection volumes roughly an order of magnitude larger. Combined with extensive structural and geophysical constraints, this history allows us to interrogate the long-term geomechanical impacts of reservoir operations. Here, we assess (i) how the  initial stress state, typically uncertain in the shallow crust ($<5$ km depth), influences  subsidence and uplift, and (ii) how production and injection operations affect fault stability. Our numerical model, calibrated with published measurements of reservoir pressures and surface displacements, incorporates a detailed representation of fault surfaces within and around the field, well-level production and injection schedules, and an elastoplastic constitutive framework. Model results show that the previously assumed stress regime in the field (reverse faulting) needs to be reassessed---the best match to the ground deformation data is achieved when the sedimentary section is initialized with low deviatoric stress (i.e., {\em not} critically stressed). This suggests significant variation in the stress state with depth, including a likely change in the stress regime. DCFF values suggest minor destabilization on reservoir faults and larger changes on sub-horizontal bedding planes; both could explain the faulting that led to sheared wells and seismicity between 1947 and 1961.
\end{abstract}

\section*{Plain Language Summary}
    Fluid injection and extraction in subsurface reservoirs can modify the stress state and impact seismicity. This is particularly important in earthquake-prone regions, such as California. The Wilmington Oil Field is the largest in the Los Angeles Basin and has produced oil since the 1930s. Early production caused nearly 9 m of ground subsidence, prompting large-scale water injection to stabilize the surface. Because the field has a long operating history and is well characterized, it is a useful setting to examine how decades of reservoir operations can affect ground movement and fault stability.

    We developed a 3D model that couples reservoir fluid flow with rock deformation as pressures change. The model includes major faults and well-by-well production and injection schedules, and was validated against published leveling measurements. Our results indicate that the reservoir was initially under less horizontal compression than previously assumed. Additionally, production-related stress changes increased the tendency for slip along shallow, sub-horizontal layers, consistent with historical reports of wells being sheared between 1947 and 1961.

    These findings highlight that initial stress conditions, typically uncertain in the shallow crust, strongly influence predicted ground displacement and fault slip. This has direct relevance for safe subsurface energy production and storage.

\section{Introduction}

    Energy-related operations in subsurface reservoirs involve fluid extraction and/or injection. This alters pore pressure and effective stress, and thereby controls the mechanical response of the subsurface~\cite{terzaghi1943,hubbert1959}. The resulting geomechanical hazards include ground deformation and fault reactivation. Early field observations linked fluid extraction to surface subsidence, including cases at Goose Creek, Texas~\cite{pratt1926}, and later at Wilmington, California~\cite{allen1970}. Other foundational studies established the connection between fluid injection, elevated pore pressure, and earthquake occurrence, including at the Rocky Mountain Arsenal disposal well near Denver~\cite{evans1966} and the controlled injection experiments at Rangely, Colorado~\cite{raleigh1976}. Induced earthquakes, in particular, have risen to prominence in the last two decades due to the challenges they continue to pose for the safe development and expansion of subsurface energy resources~\cite{ellsworth2013,hager2021,krevor2023,horne2025}.

    Reservoir production via pressure depletion leads to stress changes and surface subsidence, a relationship captured by seminal studies of poroelastic deformation~\cite{geertsma1973,segall1989}. Subsidence can often be mitigated by maintaining reservoir pressure through water injection~\cite{pierce1970}; however, field observations show that subsidence may persist despite sustained injection, reflecting inelastic compaction or fault-related deformation, as discussed for Wilmington~\cite{kosloff1980a,kosloff1980b} and the Valhall and Ekofisk fields \cite{zoback2002}. Beyond ground deformation, reservoir operations can also promote fault slip; fault reactivation may be seismic or aseismic: injection can drive slow slip/creep that can shear wells, redistribute stresses, transiently modify fault-zone permeability, and, in some cases, may precede or promote seismic rupture \cite{guglielmi2015,grigoli2017}, which is often the more safety-critical response. In particular, widespread wastewater disposal associated with oil and gas development over the last two decades has highlighted the link between fluid injection and induced seismicity~\cite{ellsworth2013}. Building on foundational demonstrations at Rangely~\cite{raleigh1976,byrne2020,silva2021}, recent field experiments show how injection can trigger fault slip (including slow slip) and transiently modify fault-zone permeability and friction~\cite{guglielmi2015,cappa2022}. Such mechanistic understanding is key in recent process-based approaches that integrate subsurface characterization, operational data, and monitoring to mitigate induced seismicity~\cite{hager2021}. This motivates coupled flow--geomechanics assessments for emerging subsurface energy technologies~\cite{viswanathan2022,krevor2023}, including geologic CO$_2$ sequestration~\cite<e.g.,>{cappa2011,jha2014,meguerdijian2021,tang2022,silva2023,lu2025} and enhanced geothermal systems, where induced earthquakes remain a key challenge~\cite<e.g.,>{parisio2019,boyet2025,horne2025}.
    
    Ongoing subsurface energy production in southern California began in the late 19th century, making the Los Angeles Basin among the most extensively developed petroleum provinces in the world~\cite{dog1964,doggr1992}. This long operational history, combined with high population density (greater Los Angeles is home to more than 18 million people) and underlying and adjacent active fault systems~\cite{wolfe2019,toghramadjian2024,plesch2026}, makes the region an important setting for evaluating the impact of decades-long subsurface energy operations on the state of stress, ground motions, and seismicity. The risks (and societal consequences) of subsurface operations are highlighted by incidents including the aforementioned production-induced subsidence and slip events at the Wilmington Field~\cite{allen1970,pierce1970,kovach1974}. In parallel, concerns about potentially induced seismicity in southern California have motivated recent studies on how reservoir operations may have affected earthquake occurrence~\cite{hauksson2015,hough2016,hough2018}. 

    Wilmington, as the largest oil field in the basin, provides a uniquely rich operational, geologic, and geophysical record, making it an attractive setting for evaluating these effects using detailed coupled flow and geomechanics models. Discovery and initial development of the Wilmington Oil Field date back to the 1930s, and subsequent production and injection led to one of the best-known examples of long-term, production-induced ground deformation (see \S~\ref{sec:data} for details). The magnitude and societal impacts of subsidence motivated foundational studies of reservoir compaction and rebound~\cite{allen1970,pierce1970,kosloff1980a,kosloff1980b}. Previous work also reported bedding-plane slip leading to well shearing in the mid-20th century~\cite{kovach1974}. Despite this history, key questions remain regarding the impacts of geologic structure and background stress state on the geomechanical response. These gaps are relevant beyond Wilmington, because predicted patterns and magnitudes of surface deformation and fault destabilization can vary strongly with assumptions about fault geometry and initial stress conditions.

    Here, we seek to advance our understanding of the initial faulting regime (also referred to as stress state or background stress regime), the changes in stress state due to subsurface energy operations, and how these influenced ground deformation and fault stability at Wilmington. We develop a 3D coupled multiphase flow--geomechanics model that incorporates a detailed subsurface description, including 24 faults within and around the main producing interval, together with well-level production and injection schedules (\S~\ref{sec:methods}). To capture permanent deformation, we employ an elastoplastic constitutive framework. We constrain the model using published reservoir pressure data (\S~\ref{sec:results_reservoir}) and ground subsidence/rebound from leveling surveys (\S~\ref{sec:results_mech}). We then assess the impact of the background stress regime (\S~\ref{sec:results_mech}), and quantify stability changes on faults and sub-horizontal surfaces (\S~\ref{sec:fault_stability}).

\section{Historical data} \label{sec:data}
    \subsection{Field characteristics and reservoir operations} \label{sec:field}
\begin{figure}[ht]    
                \centering
                \includegraphics[width=1.0\textwidth]{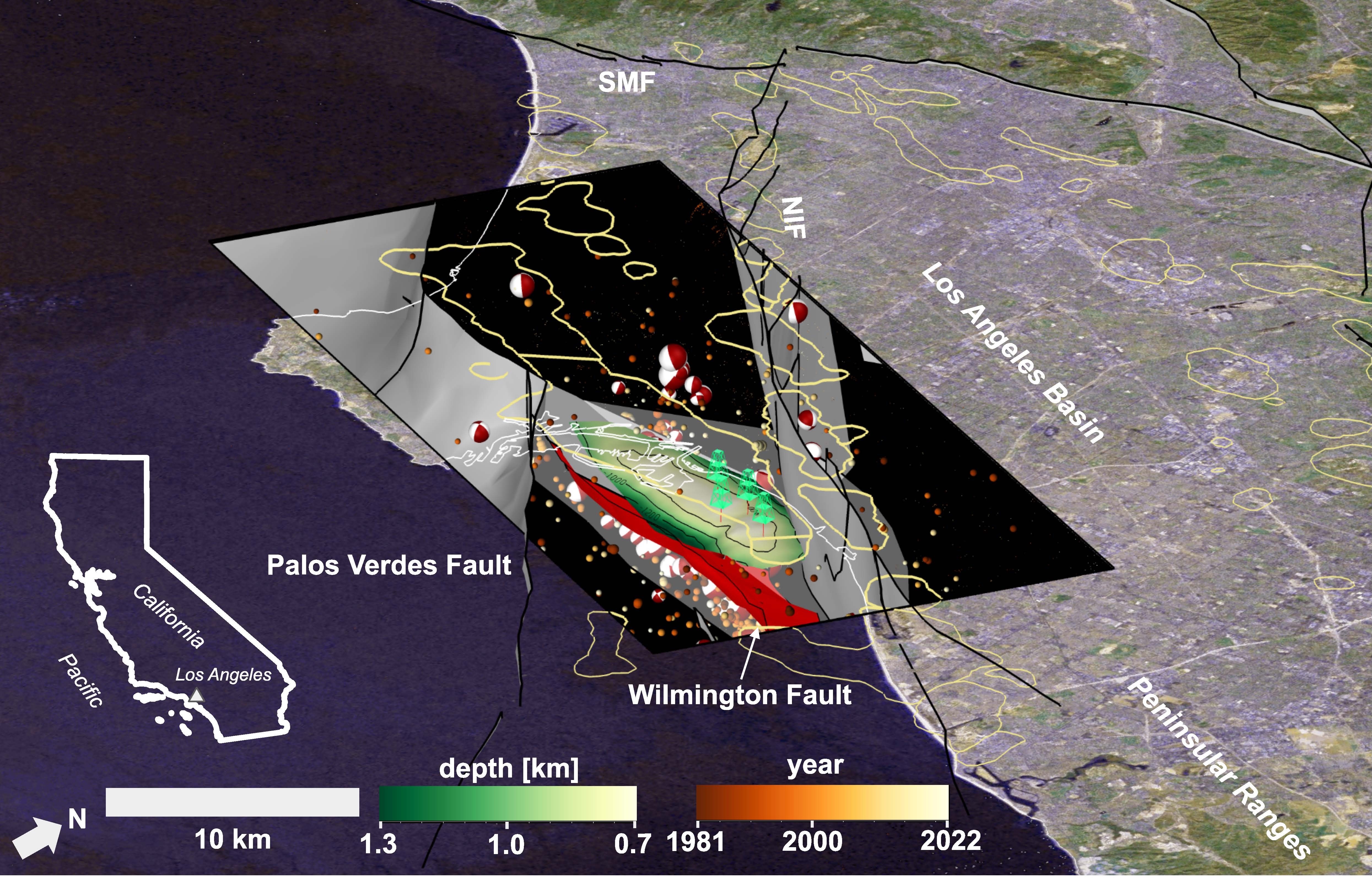}
                \caption{Perspective view of the study area (see inset for location) and the subsurface context of the Wilmington field (window cutout). Green color scale and contours: depth of reservoir horizon~\cite{dog1964,doggr1992}; green well derricks: active well platforms; shaded red and grey: 3D fault surfaces \cite{plesch2026}; brown color scale: hypocenters colored by year of occurrence \cite[and updates]{hauksson2012}; quartered spheres: earthquake focal mechanisms \cite[and updates; compressive quadrants maroon]{yang2012}; black lines: fault traces; yellow lines: hydrocarbon fields; NIF: Newport Inglewood Fault; SMF: Santa Monica Fault.}
                \label{fig:basin}
        \end{figure}

\begin{figure}[h!]    
            \centering
            \includegraphics[width=1.0\textwidth]{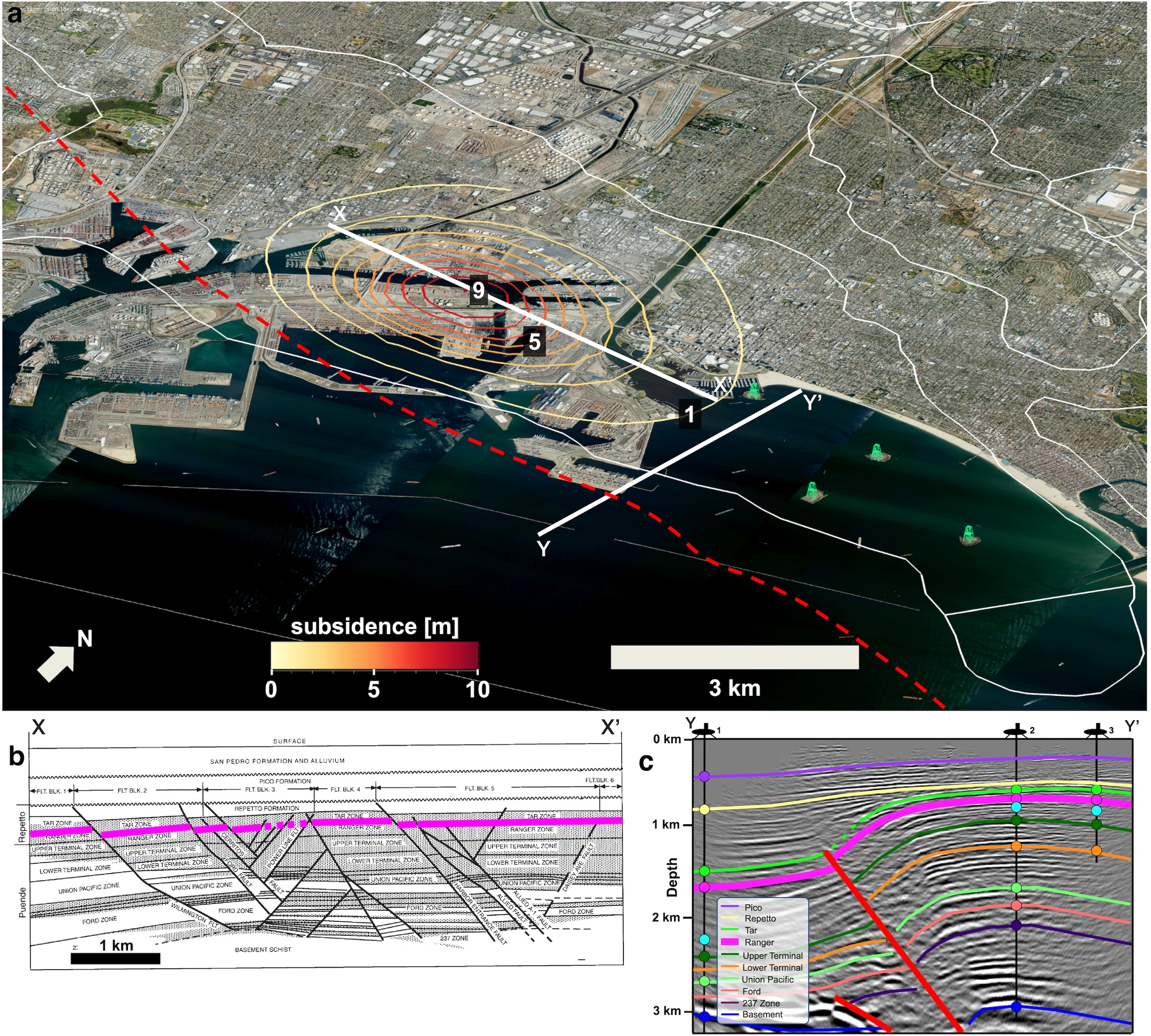}
            \caption{\textbf{a}: Perspective view of total subsidence in the Wilmington field~\cite<color contours, 1~m interval, from>{mayuga1970}. White outlines: field boundaries; green well derricks: THUMS oil islands; stippled red line: tip line of blind Wilmington thrust~\cite{wolfe2019,plesch2026}; white lines: location of cross-sections; imagery: NAIP (USDA Farm Services Agency). \textbf{b}: cross-section X--X$^\prime$~\cite<modified from>{montgomery1998}; no vertical exaggeration; bold purple line: top of Ranger zone. \textbf{c}:  cross-section Y--Y$^\prime$~\cite<modified from>{wolfe2019}; no vertical exaggeration; bold purple line: top of Ranger zone; red lines: blind Wilmington thrust; dip-meter data (yellow) and horizon picks (colored circles) from various THUMS oil islands (locations 2 and 3) and BETA deviated oil wells (location 1) are shown. Not all boreholes are shown for clarity. Seismic reflection survey source: Long Beach 3D Seismic Survey, January 1995~\cite{otott1996}. See Open Research section.} 
             \label{fig:field_data}
        \end{figure}

        The Wilmington Oil Field is located in the Los Angeles Basin (Fig.~\ref{fig:basin}), which formed during Miocene oblique rifting of the continental margin~\cite{luyendyk1985,crouch1993} and is filled primarily by Miocene and Pliocene siliciclastic sediments~\cite{mayuga1970,blake1991}. The Puente Fm (Miocene) and the Repetto Fm (Pliocene) host the Wilmington reservoirs, located in seven productive zones. From bottom to top, these zones are named 237, Ford, Union Pacific, Lower Terminal, Upper Terminal, Ranger, and Tar~\cite{montgomery1998}(Fig.~\ref{fig:field_data}). The Ranger Zone is the most prolific, and is estimated to account for $\approx~50$\% of the original oil in place~\cite{kwong2019}. Structurally, the Wilmington Field is located in a northwest-trending anticline between the Palos Verdes (to the West) and the Newport-Inglewood (to the East) strike-slip fault zones. This anticline was formed due to displacement along the Wilmington Fault, a Miocene normal fault reactivated in the Pliocene with reverse slip~\cite{wolfe2019}. Multiple transverse normal faults, most of which have been reported to be sealing, divide the field into several fault blocks~\cite{montgomery1998}(Fig.~\ref{fig:field_data}b).

        The field was discovered in 1932, with formal identification as a separate field from the adjacent Torrance Field in 1936, and rapid development afterwards~\cite{truex1969,otott1996}. Production from the Tar, Ranger, and Upper Terminal was typically from wells completed through all three zones, with the Ranger and Upper Terminal Zones occurring over a depth interval of $\approx$250 m~\cite{nelson1976}. Primary oil production was through solution gas drive, which depleted the reservoir pressure by more than 80 bar during the first 25 y of production~\cite{dog1964}. Waterflooding operations were initiated around 1958 after several pilots~\cite{pierce1970}. Successive secondary recovery approaches were implemented to enhance production of bypassed oil, and the average field pressures recovered to hydrostatic levels. Cumulative oil production to present day is $\approx2.5$ billion barrels (BBO)~\cite{kwong2019}. Detailed production figures based on well data are presented in \S~\ref{sec:results_reservoir}. 

        Table~\ref{tab:zones} shows the initial petrophysical and oil characteristics across the production zones. The Ranger Zone displays high porosity and permeability, moderate brine salinity, and contained viscous (80 cP at 59$^\circ$C) heavy to medium (12–25 $^\circ$API) crude oil. Initial oil saturations are relatively high, while both water and gas saturations are comparatively low, indicating a favorable reservoir for oil recovery. Similar characteristics, but with differences in oil viscosity, are observed in the Upper Terminal and Tar Zones, which together with Ranger contain more than 75\% of the oil reserves estimated at Wilmington. 

        \begin{table}[htbp]
            \centering
            \caption{Reservoir and fluids data for the different productive zones at Wilmington. Data compiled for this table is from~\citeA{mayuga1970,doggr1992,montgomery1998}. Single number indicates average if followed by $^*$, GOR $=$ solution gas-oil ratio, $S_{\text{wi}}=$ initial (irreducible) water saturation, $S_{\text{oi}}=$ initial (maximum) oil saturation, $S_{\text{gi}}$ initial gas saturation.}
            \label{tab:zones}
            {\footnotesize
            \begin{tabular}{lccccccc}
                \hline
                & \multicolumn{7}{c}{Production zones} \\
                \cline{2-8}
                Properties     & 237 & Ford & Union Pacific & Lo. Terminal & Up. Terminal & 
                Ranger & Tar \\
                \hline
                Thickness [m]         & 61–366 & 229–366 & 122–274 & 152–244 & 122–259 & 122–229 & 76–229 \\
                Sand [\%]             & 40$^*$ & 25-35 & 20-25 & 50-70 & 50-70 & 30-35 & 35-40 \\
                Porosity [\%]         & 25$^*$ & 25$^*$ & 20-25 & 25-35 & 25-37 & 27-37 & 30-40 \\
                Permeability [mD]     & 275$^*$ & 100$^*$ & 150$^*$ & 450$^*$ & 700$^*$ & 700-1500 & $>$1000$^*$ \\
                Temperature [$^\circ$C]& 109 & 98 & 82 & 74 & 66 & 63 & 51 \\
                Oil gravity [$^\circ$API] & 28-32 & 28-32 & 25-32 & 20-31 & 14-25 & 12-25 & 12-15\\
                GOR [scf/stb]         & 542 & 463-473 & 322-338 & 215-310 & 156-207 & 140-160 & 94-100 \\
                Oil viscosity [cP] (T)&  & 105 (109) &  & 40 (73) & 45 (60) & 80 (59) & 283 (52) \\
                Brine salinity [ppt]  & 28 & 31 & 32-34 & 30-32 & 33 & 32 & 21-28 \\
                $S_{\text{wi}}$ [-]   & 0.32 & 0.47 & 0.42-0.46 & 0.36-0.39 & 0.3-0.33 & 0.25-0.35 & 0.22-0.31 \\
                $S_{\text{oi}}$ [-]   & 0.68 & 0.53 & 0.54-0.58 & 0.61 & 0.67 & 0.62-0.75 & 0.64-0.78 \\
                $S_{\text{gi}}$ [-]   & 0 & 0 & 0 & 0-0.03 & 0-0.03 & 0-0.03 & 0-0.05 \\
                \hline
            \end{tabular}
            }
        \end{table}

    \subsection{Ground deformation} \label{sec:ground_def}
        Subsidence began shortly after the start of production and continued until $\approx$1965~\cite{mayuga1970}. Surveys conducted by the US Coast and Geodetic Survey in 1941, 1947, 1951, 1954, 1958, and 1965 showed a maximum subsidence of 0.4 m (1.3 ft), 2.4 m (7.9 ft), 4 m (13.1 ft), 6.1 m (20 ft), 7.9 m (25.9 ft), and 8.8 m (29 ft),  respectively~\cite{kovach1974,kosloff1980a}. Maximum horizontal displacements of 3.7 m (12.1 ft) were also recorded, and the area of the subsidence bowl reached 50 km$^2$ (20 mi$^2$)(Fig~\ref{fig:field_data}). Waterflooding operations, started in the 1950s, effectively stopped subsidence by 1968. Despite the fact that reservoir pressures were brought back to pre-production levels by 1970, the maximum observed uplift between 1958 and 1975 was only 0.4--0.5 m (1.3--1.6 ft)~\cite{kosloff1980b}. This observation demonstrates the inelastic nature and irreversibility of the subsidence process, which involved plastic compaction of the producing sands and interbedded fine-grained materials~\cite{mayuga1970,kosloff1980b}. Subsidence time-series figures based on surface measurements are presented in \S~\ref{sec:results_mech}. 
        
    \subsection{Stress state and seismicity} \label{sec:stress}
         Los Angeles Basin is located to the SE of the Santa Monica Fault and NW of the Peninsular Ranges (Fig.~\ref{fig:basin}). Active, transpressive deformation is partitioned into NW-SE and E-W trending strike-slip fault and blind-thrust systems based on extensive geologic, geodetic, and seismologic observations~\cite{dolan1995,argus1999,shaw2002}. The Wilmington field is generally considered to lie within a reverse faulting regime~\cite{yang2013stress,wolfe2019}, bounded to the northeast and southwest by the strike-slip Newport-Inglewood and Palos Verdes fault zones, respectively. Regionally, the orientation of the maximum horizontal stress ($\sigma_\text{H}$) remains at a N to NE orientation in the upper 15 km of the crust, based on data from hydraulic fracturing, borehole breakouts, and focal mechanism inversion. The orientation of the minimum horizontal stress ($\sigma_\text{h}$) is also generally uniform (W to NW)~\cite{zoback1980,hauksson1990,mount1992,lundstern2020,luttrell2021}. If the principal stress axes are consistently oriented, the strike- and reverse-slip in the basin suggests that either two of the principal stresses have similar magnitude, or their relative magnitudes vary significantly with depth~\cite{zoback1980}. In the LA basin, both are locally present: Significant heterogeneity exists with depth between the shallow and deep portions of the sedimentary cover, and also laterally near the basin edges; the stress orientations and regime in the basement are more homogeneous, with predominantly reverse faulting at 10-15 km depth; and quantification of the relative stress magnitudes by means of~\citeA{simpson1997} $A_\phi$ parameter (see Eq~\ref{eq:aphi}) is $\approx [1.6, 2.1]$, consistent with strike-slip/reverse faulting regime, depending on the model and depth~\cite{hardebeck2001,yang2013stress,lundstern2020,luttrell2021,luttrell2025}.  
         
         \citeA{hauksson1990} evaluated focal mechanisms for 244 events with magnitudes of at least 2.5 that occurred between 1977 and 1989. 59\% were strike-slip and mostly located near the Newport-Inglewood and Palos Verdes Fault Zones; 32\% had reverse mechanisms, occurring below 4--6 km depth along the Elysian Park and Torrance--Wilmington fold and thrust belts; and 9\% were normal faulting events. In the eastern part of the basin, \citeA{yang2011} obtained similar percentages and showed that normal faulting has the highest probability density at shallow depths (0--6 km), while strike-slip and reverse faulting styles have higher probability densities at greater depth. In general, the coexistence of both strike-slip and thrust mechanisms indicates partitioning of slip, rather than oblique slip. Basin-wide, almost all seismicity occurs deeper than 5 km, in the basement (see Fig.~\ref{fig:basin}), while hydrocarbon wells rarely reach below 5 km~\cite{hauksson2015}. 
         
         At Wilmington, shallow M$_\text{L} \approx 3$ earthquakes occurred in 1947, 1949, 1951, 1954, 1955, and 1961. \citeA{kovach1974} argued that subsidence increased shear stresses on the flanks of the subsidence bowl, which were released via seismic slip on two main planes; these slip planes were thin ($\approx$2 m) shale beds located at a depth of $\approx$500 m. Cores recovered from the field showed slickensides where most well damage was observed, but focal mechanisms have not been reported. These earthquakes have been referred to as ``shallow slumping events''~\cite{hough2018}. Basin-wide, \citeA{hauksson2015} found no significant differences between seismicity occurring inside and outside oil fields, or correlation with fluid injection/production since 1977; they concluded that there is no clear evidence of induced earthquakes. In contrast, \citeA{hough2018} focused on seven independent $M_\text{L} \geq 4$ events occurred between 1932 and 1952. They used macroseismic data to revise early earthquake locations and modeled Coulomb static stress changes due to rectangular deflating sources representing oil production in eight fields, including Wilmington. Although they note that focal mechanisms are unkonwn and hypocentral depths poorly constrained, they proposed epicenters that are near the boundaries of active fields, and they found stress changes $>10$ bar close to the oil fields. They concluded that a number of significant earthquakes between 1900 and 1950 may have been induced~\cite<see also>{hough2016}.

\section{Methods} \label{sec:methods}
    \subsection{Coupled multiphase flow and geomechanics modeling}
        Here, we use a one-way coupled flow--geomechanics modeling framework~\cite<e.g.,>{kosloff1980a, kosloff1980b, silva2021, silva2023}. In this framework, we first solve for pore pressures and saturations throughout the production history (1936-2020). The changes in pore pressure at each time step are then used as internal volumetric loads to drive geomechanical deformation (\S ~\ref{sec:numerical}). This approach allows us to quantify effective stress changes in the computational domain that result from direct pore pressure variations as well as mechanical deformation. The validity of the one-way coupling will be justified in the results, where we demonstrate that observed ground deformations can be reproduced with our model. In this section, we present the governing equations and numerical solution approach; the chosen parameter values are given in \S~\ref{sec:setup}.

        \subsubsection{Governing equations for multiphase flow}
            Hydrocarbon recovery and water injection at Wilmington can be modeled using the industry-standard black-oil formulation~\cite{aziz1979,lie2019}. In this formulation, each fluid phase (aqueous, oleic, and gaseous) is primarily composed of a single pseudo-component with the corresponding name (water/brine, oil, and gas). The mass conservation equations for each component $l$ are written as follows:
            \begin{equation}
                \begin{aligned} \label{eq:black-oil}
                    \partial_t (\phi b_w S_w ) + \nabla \cdot (b_w \mathbf{q}_w) = Q_w\\[2mm]
                    \partial_t (\phi b_o S_o ) + \nabla \cdot (b_o \mathbf{q}_o) = Q_o \\[2mm]
                    \partial_t [\phi(b_g S_g + R_s b_o S_o)] + \nabla \cdot (b_g \mathbf{q}_g + R_s b_o \mathbf{q}_o) = Q_g
                \end{aligned}
            \end{equation}
            where $\phi$ is porosity, $S$ saturation, and $Q$ denotes source terms. As written, Eq.~\ref{eq:black-oil} considers that the water and oil components are only present in their respective phases, whereas the gas component can also be dissolved in the oil phase. Dissolution of natural gas in the oil as a function of pressure is accounted for through the gas-oil solution ratio, which is defined as $R_s (p) = V_g^s / V_o^s$. Here, the superscript $s$ indicates stock-tank conditions, defined at 15.56 $^\circ$C and 1 atm. In Eq.~\ref{eq:black-oil} we also used shrinkage factors~\cite{lie2019}, defined for each component as $b_l = V_l^s / V_l$, i.e., the reciprocal of the standard formation volume factors. The volumetric flux of each phase ($\mathbf{q}_\alpha$) is modeled via Darcy's law for multiphase flow~\cite{muskat1949}:           
            \begin{equation} \label{eq:darcy}
                \mathbf{q}_\alpha = - \frac{k k_{r\alpha}}{\mu_\alpha}(\nabla p_{\text{f},\alpha} - \rho_\alpha \mathbf{g}).
            \end{equation}
            where $k$ is the intrinsic permeability, $k_{r}$ is the relative permeability, $\mu$ the dynamic viscosity, $p_\text{f}$ the fluid pressure, $\rho$ the mass density, and $\mathbf{g}$ the gravity vector. As discussed in \S~\ref{sec:setup}, we consider intrinsic permeability to be a scalar and capillary pressure to be 0; this is because fluid-flow is only modeled in the main producing interval (Ranger formation). Relative permeability is modeled following the ECLIPSE standard model~\cite{schlumberger2014technical}, in which the relative permeability of water and gas are taken directly from two-phase input data (see \S~\ref{sec:setup}), and the oil relative permeability is linearly interpolated:
            \begin{equation}
                \begin{aligned}
                    k_{rw}(S_w)              & = k_{rw}^{wo}(S_w) \\[2mm]
                    k_{ro}(S_w, S_g)         & = \frac{S_g k_{ro}^{og}(S_o)}{S_g + S_w - S_{wc}}
                                               + \frac{(S_w - S_{wc}) k_{ro}^{wo}(S_o)}{S_g + S_w - S_{wc}} \\[2mm]
                    k_{rg}(S_g)              & = k_{rg}^{og}(S_g)
                \end{aligned}
                \label{eq:kr}
            \end{equation}
            where $S_o = 1 - S_w - S_g$ and $S_{wc}$ is the connate or irreducible water saturation. This equation for the oil relative permeability is based on the Stone I model~\cite{stone1961}, proposed for water-wet porous media, but uses a linear interpolation to improve numerical convergence~\cite<see discussion in>[\S11.3]{lie2019}.

        \subsubsection{Governing equations for poromechanics} \label{sec:poromech}
            The conservation law for mechanics is the balance of linear momentum~\cite<e.g.>{wang2000}, written here in the quasi-static form neglecting inertial terms:
            \begin{equation} \label{eq:balance}
                \nabla \cdot (\boldsymbol{\sigma}^\prime - bp_\text{f}\mathbf{1}) + \rho_\text{b} \mathbf{g} = \mathbf{0}
            \end{equation}
            where $\rho_b$ is the bulk density. In a porous medium, deformation is controlled by the effective stress, $\boldsymbol{\sigma}^\prime$~\cite{terzaghi1943}:
            \begin{equation} \label{eq:effective}
                \boldsymbol{\sigma}^\prime = \boldsymbol{\sigma} + bp_\text{f}\mathbf{1}
            \end{equation}
            where $\boldsymbol{\sigma}$ is the Cauchy total stress tensor and $b$ Biot's coefficient~\cite{biot1941}. The sign convention adopted here is such that tension is positive. Thus, in Eq.~\ref{eq:balance}, the total stress is expressed using Eq.~\ref{eq:effective} as a function of the effective stress and the pore pressure. 

            We assume that the deformation behavior observed in the field can be reproduced using an elastoplastic constitutive model. In the framework adopted here, we use a finite-strain, incremental plasticity theory with additive strain-rate decomposition~\cite{dunne2005,borja2013}:
            \begin{equation}
                \dot{\boldsymbol{\varepsilon}} = \dot{\boldsymbol{\varepsilon}}^\text{el} + \dot{\boldsymbol{\varepsilon}}^\text{pl}
            \end{equation}
            where the superscripts $\text{el}$ and $\text{pl}$ refer to the elastic and plastic parts, respectively. For the elastic response, we assume linear isotropic elasticity, such that the effective stress rate is linearly related to the elastic strain rate by
            \begin{equation} \label{eq:elasticity}
                \dot{\boldsymbol{\sigma}}^\prime = \mathbb{C} : \dot{\boldsymbol{\varepsilon}}^\text{el}
            \end{equation}
            where $\mathbb{C}$ is the fourth-order elasticity tensor, which in this case depends only on the Young's modulus ($E$) and Poisson's ratio ($\nu$), which remain constant in our model. We emphasize that, in Eq.~\ref{eq:elasticity}, the computed stresses are {\em effective} stresses, in agreement with the concept introduced in Eq.~\ref{eq:effective}. 

\begin{figure}[h]    
                \centering
                \includegraphics[width=1.0\textwidth]{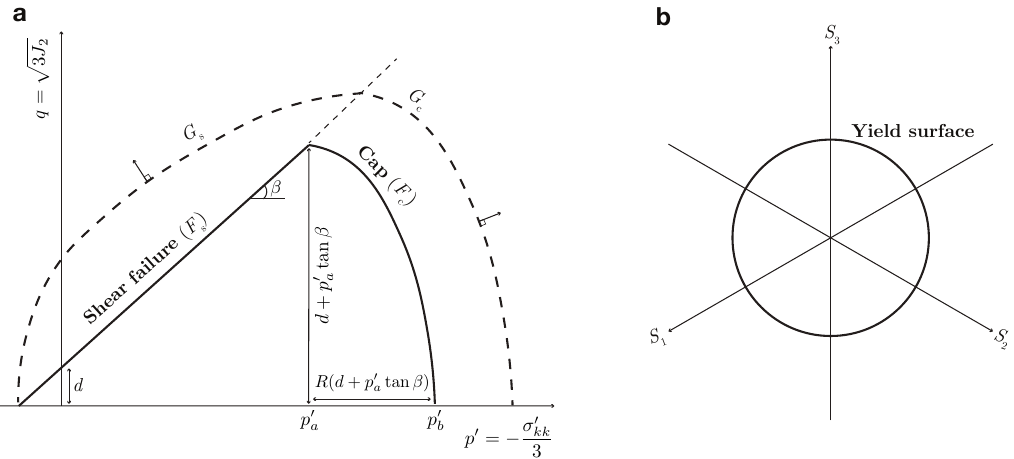}
                \caption{\textbf{a}: Modified Drucker-Prager Cap Model in the $p^\prime$ - $q$ plane, where $p^\prime$ is the mean effective stress and $q$ is the von Mises stress. The model is defined by a shear failure envelope closed by a cap. The flow potential surface, shown with dashed lines, has a part identical to the cap yield surface ($G_\text{c}$) and another elliptical part in the shear failure region ($G_\text{s}$). \textbf{b} The yield surface shown in the $\pi$-plane, defined here as a von Mises circle with radius depending on $p^\prime$.}
                \label{fig:dpcm}
            \end{figure}

            For the plastic response, we employed a modified Drucker-Prager Cap Model (DPCM). The model is defined in terms of the scalar stress measures $p^\prime$ and $q$ (Fig.~\ref{fig:dpcm})~\cite{puzrin2012,borja2013,abaqus2023}:
            \begin{equation}
                \begin{aligned}
                    p^\prime &= -\frac{I^\prime_1}{3} = -\frac{1}{3} \sum_k \sigma^\prime_{kk}\\[2mm]
                    q &= \sqrt{3 J_2} = \sqrt{\frac{3}{2} (S_1^2 + S_2^2 + S_3^2)}\\
                \end{aligned}
                \label{eq:pq}
            \end{equation}
            where $I^\prime_1$ is the first invariant of the effective stress tensor, and $J_2$ the second invariant of the deviatoric stress tensor ($\boldsymbol{S}$) defined as $\boldsymbol{S} = \boldsymbol{\sigma^\prime} + p^\prime\boldsymbol{1}$. The shear failure and cap yield surfaces, the latter allowing us to model plastic compaction, are:
            \begin{equation}
                \begin{aligned}
                    F_\text{s} &= q - p^\prime \tan \beta - d = 0\\[2mm]
                    F_\text{c} &= \sqrt{(p^\prime-p^\prime_a)^2 + (Rq)^2} - R(d + p^\prime_a \tan \beta) = 0\\
                \end{aligned}
                \label{eq:f}
            \end{equation}
            where $\tan \beta = \frac{6\sin\phi^\prime}{3 - \sin\phi^\prime}$ is the slope in $p^\prime$-$q$ plane, related to the effective friction angle of the material ($\phi^\prime$); $d = \frac{6 c^\prime \cos \phi^\prime}{3 - \sin\phi^\prime}$ is the Drucker-Prager cohesion parameter; $R$ is the cap eccentricity parameter; $p^\prime_b$ is the hydrostatic compression yield stress, modeled as a function of the plastic strain (\S~\ref{sec:setup}); and the evolution parameter $p^\prime_a$ is:
            \begin{equation} \label{eq:pa}
               p^\prime_a = \frac{p^\prime_b - Rd}{1 + R \tan \beta}
            \end{equation}
            The plastic flow potential is modeled by two elliptic sections ($G_\text{s}$, $G_\text{c}$), the latter remaining the same as in $F_\text{c}$:
            \begin{equation}
                \begin{aligned}
                    G_\text{s} &= \sqrt{[(p^\prime_a - p^\prime)\tan \beta]^2 + q^2}\\[2mm]
                    G_\text{c} &= \sqrt{(p^\prime - p^\prime_a)^2 + (Rq)^2}\\
                \end{aligned}
                \label{eq:g}
            \end{equation}
            Hence, the plastic flow potential is associated in the cap, and non-associated in the Drucker-Prager failure surface.            

        \subsubsection{Numerical solution approach} \label{sec:numerical}
            The multiphase flow problem is solved first using the Matlab Reservoir Simulation Toolbox~(\texttt{MRST}), an open-source, finite-volume-based toolbox for simulation of multiphase flows in the subsurface~\cite{lie2019,lie2021}. The isothermal \texttt{ad-blackoil} module of \texttt{MRST}, used here, employs a fully implicit temporal discretization and the industry-standard two-point flux approximation for transmissibility calculation \footnote{\url{https://www.sintef.no/projectweb/mrst/modules/ad-core/}} \cite<e.g.>{landa2021,silva2023,salo2024mrst}. According to Eq.~\ref{eq:black-oil}, we consider dissolution/exsolution of the natural gas in the oleic phase, while the aqueous and gaseous phases are fully composed of their respective components.

            The mechanics problem is solved once the pore pressures are known at all spatial locations and timesteps. To solve the mechanics problem driven by pore pressure changes, we used the commercial finite-element simulator \texttt{Abaqus}~\cite{abaqus2023}, widely used to model hydro-mechanical processes~\cite<e.g.>{bisdom2016,hager2021,haddad2023,sun2023,jin2024}. For the constitutive response, we employed the \texttt{*Cap plasticity} and \texttt{*Cap hardening} keywords to apply the modified DPCM described above. The starting point for the finite element discretization is the weak form of the equilibrium statement (Eq.~\ref{eq:balance}), i.e., the virtual work equation~\cite{hughes2000,jha2014,abaqus2023}:
            \begin{equation} \label{eq:weak}
                \int_V (\boldsymbol{\sigma}^\prime - bp_\text{f}\boldsymbol{1}) : \delta \boldsymbol{D}\; dV = \int_S \mathbf{t} \cdot \delta \mathbf{v}\; dS + \int_V \mathbf{f} \cdot \delta \mathbf{v}\; dV 
            \end{equation}
            Where $\delta \boldsymbol{D} = \frac{1}{2}(\nabla \delta \mathbf{v} + (\nabla \delta \mathbf{v})^T)$ is the virtual strain rate, i.e.~the symmetric part of the gradient of the virtual velocity $\delta \mathbf{v}$ (which acts as our test function). For our specific problem, the surface tractions vanish given the boundary conditions (\S~\ref{sec:setup-geomechanics}), and the internal body force is given by $\rho_\text{b}\mathbf{g}$. The {\em known} pore pressures are moved to the RHS:
            \begin{equation} \label{eq:weak_fin}
                \int_V \boldsymbol{\sigma}^\prime : \delta \boldsymbol{D}\; dV = \int_V bp_\text{f} (\nabla \cdot \delta \mathbf{v})\; dV + \int_V \rho_\text{b}\mathbf{g} \cdot \delta \mathbf{v}\; dV
            \end{equation}
            In the pore pressure term, the operation $\mathbf{1} : \delta \boldsymbol{D}$ is replaced by $\nabla \cdot \delta \mathbf{v}$, since the double contraction of the identity tensor with the virtual strain gives the divergence of the virtual velocity. Thus, in Eq.~\ref{eq:weak_fin}, the pore pressure acts as an internal volumetric load. 

            From this weak form, the finite element system and Newton-Raphson nonlinear solution scheme are developed using an admissible set of interpolation functions. The residual vector and Newton update at iteration $k$ can be summarized as:
            \begin{equation}
                \begin{aligned}
                \mathbf{r}(\mathbf{u}) & = \int_V \boldsymbol{\sigma}^\prime (\mathbf{u}) : \delta \boldsymbol{D}\; dV -\int_V b p_\text{f} (\nabla \cdot \delta \mathbf{v})\; dV -\int_V \rho_\text{b}\mathbf{g} \cdot \delta \mathbf{v}\; dV \\[2mm]
                \mathbf{J}\,\Delta\mathbf{u}^k &= -\mathbf{r} \\[2mm]
                \mathbf{u}^{k+1} &=  \mathbf{u}^{k} + \Delta\mathbf{u}^k        
                \end{aligned}
            \end{equation}
            where $\mathbf{u} = \mathbf{u}(\mathbf{x})$ is the discrete displacement field, parametrized through the finite element interpolation, $\mathbf{u}(\mathbf{x}) = \sum_a N_a(\mathbf{x})\, \mathbf{u}_a$; $\mathbf{J}$ is the Jacobian matrix; and $\Delta\mathbf{u}^k$ is the Newton correction. Assembly details are provided in the texts cited above~\cite{hughes2000,jha2014,abaqus2023}.

    \subsection{Geostructural model and unified mesh}
\begin{figure}[h]  
            \centering
            \includegraphics[width=1\linewidth]{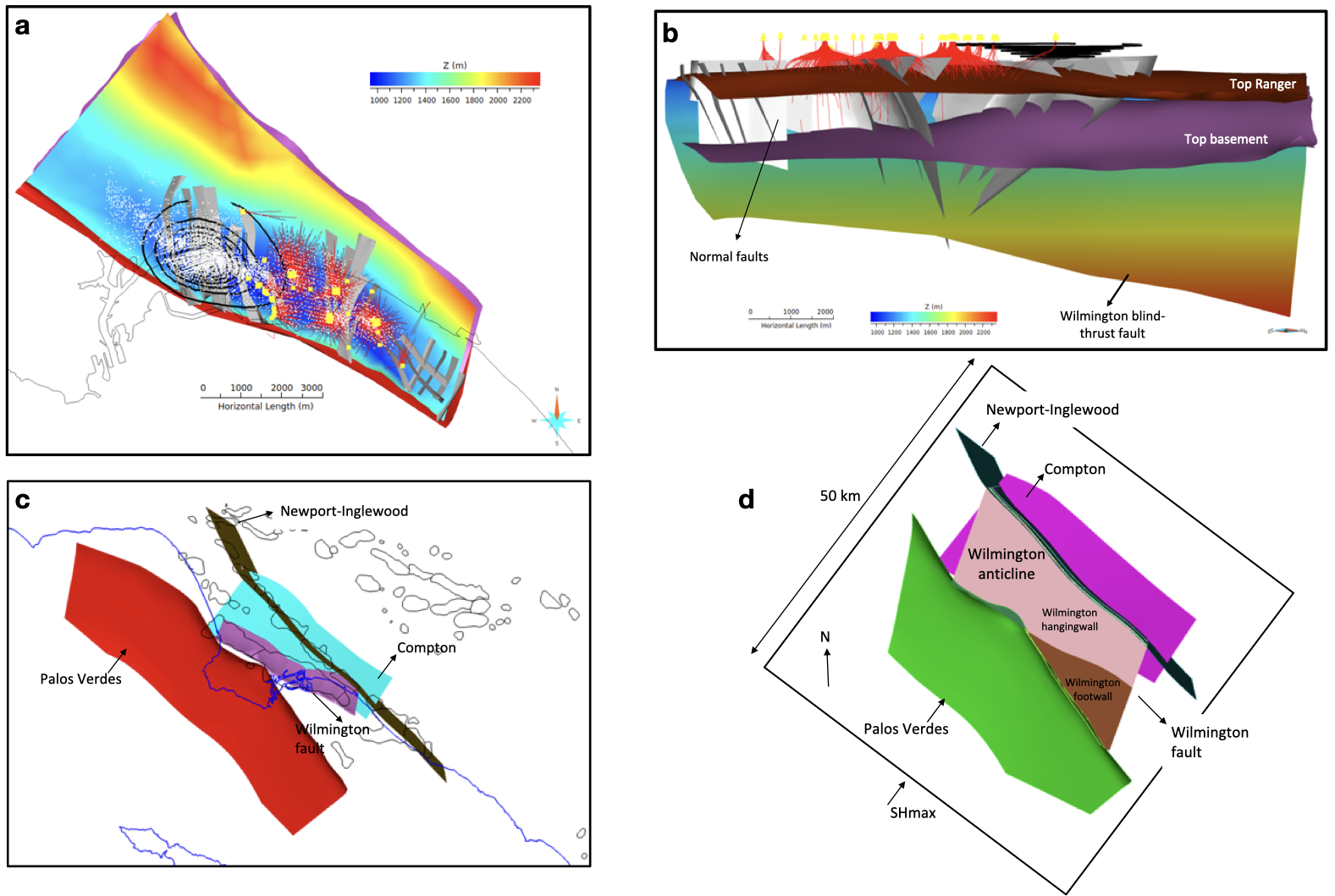}
            \caption{Geological model for the Wilmington oil field. (\textbf{a}) View from the top. Colored surface is the top of the Ranger zone (main producing interval). Black contoured lines indicate the location of the total ground subsidence between 1936 to 1955~\cite{mayuga1970}; white dots are the well locations. White surfaces are normal faults that were interpreted based on cross-sections found in the literature~\cite{wright1991}. The red surface is the Wilmington blind thrust fault. (\textbf{b}) View from the east. Red surface is the top of the Ranger interval. Purple surface is the top of the basement. The colored surface is the Wilmington fault. The white surfaces are normal faults that strike perpendicular to the Wilmington anticline. Pico seal horizon not shown here for simplicity. (\textbf{c}) Regional faults that bound the Wilmington anticline. The outline of the different oilfields in the LA basin is shown by the thin black line. (\textbf{d}) Top view of our final geological model showing the location of the regional faults with respect to the Wilmington anticline. Our model domain is 60$\times$60 km.}
            \label{fig:GeologicalModelFromGocad}
        \end{figure}

        We built a structural model for the Wilmington oil field using cross sections and oil field reservoir horizon contour maps available in the literature~\cite{wright1991, mayuga1970, montgomery1998}, as well as 2- and 3-D seismic reflection data onshore and offshore~\cite<refer to>[for references]{wolfe2019}. Our model includes a total of 24 fault segments, the most relevant of which are:
        \begin{itemize}
            \item Several northeast-southwest normal faults that cut across the Wilmington anticline, also referred to as tear faults (Fig. \ref{fig:GeologicalModelFromGocad}). These normal faults are known to be important flow barriers and serve to compartmentalize the field extensively.

            \item The Wilmington blind-thrust fault~\cite{wolfe2019}, which lies beneath the Wilmington oil field (Fig. \ref{fig:GeologicalModelFromGocad}b).

            \item The Newport--Inglewood~\cite<from>{plesch2007} and Palos Verdes Faults~\cite<from>{wolfe2022}, which are both included in the Community Fault Model for California~\cite{plesch2026}.
            
        \end{itemize}
        
        The complex stratigraphic section of the Wilmington oil field was simplified in our model to include only the Pico Fm. (seal), top and bottom Ranger (main producing zone), and the top basement. 

\begin{figure}[h]  
            \centering
            \includegraphics[width=1\linewidth]{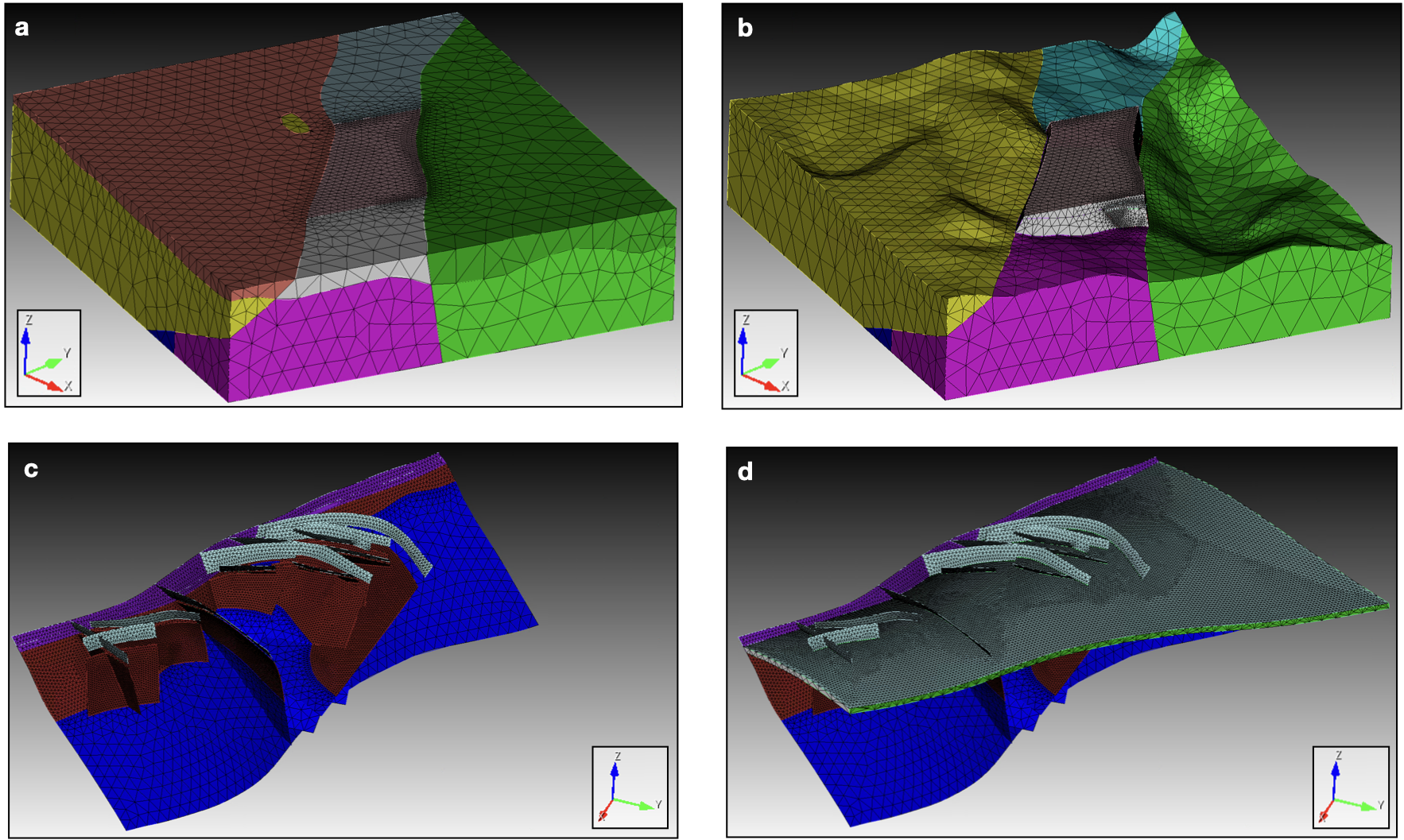}
            \caption{Overview of computational mesh. (\textbf{a}) The Wilmington oil field, located in the center of the domain, and regional faults. (\textbf{b}) View showing into the model domain, emphasizing the location of the Wilmington oil field in the center. (\textbf{c}) Embedded faults discretized in the computational mesh. (\textbf{d}) View showing the producing interval (Ranger zone) along with the faults cutting the reservoir. The east-west, north-south, and vertical directions are aligned with the X, Y, and Z directions, respectively, and the grid has dimensions of 60$\times$60$\times$15 km.}
            \label{fig:ComputationalMeshFromCubit}
        \end{figure}

        We then discretized the geostructural model using tetrahedral elements to build a unified computational mesh for both flow and geomechanical modeling (Fig.~\ref{fig:ComputationalMeshFromCubit}). The use of tetrahedral elements allows the mesh to conform to the complex geometry of the embedded faults, as well as the general geometry of the producing reservoir, which has the benefit of explicitly accounting for the impact of faults on flow modeling. The computational mesh has higher resolution within the Wilmington oil field, where the element size is on the order of ten meters; element size progressively increases towards the boundaries of the domain---especially the areas outside of the Wilmington field---where it can be hundreds of meters. The use of variable element size resulted in a computational mesh with approximately one million elements.

    \subsection{Numerical model setup} \label{sec:setup}
        \subsubsection{Reservoir simulation model}
\begin{figure}[h]    
            \centering
            \includegraphics[width=1.0\textwidth]{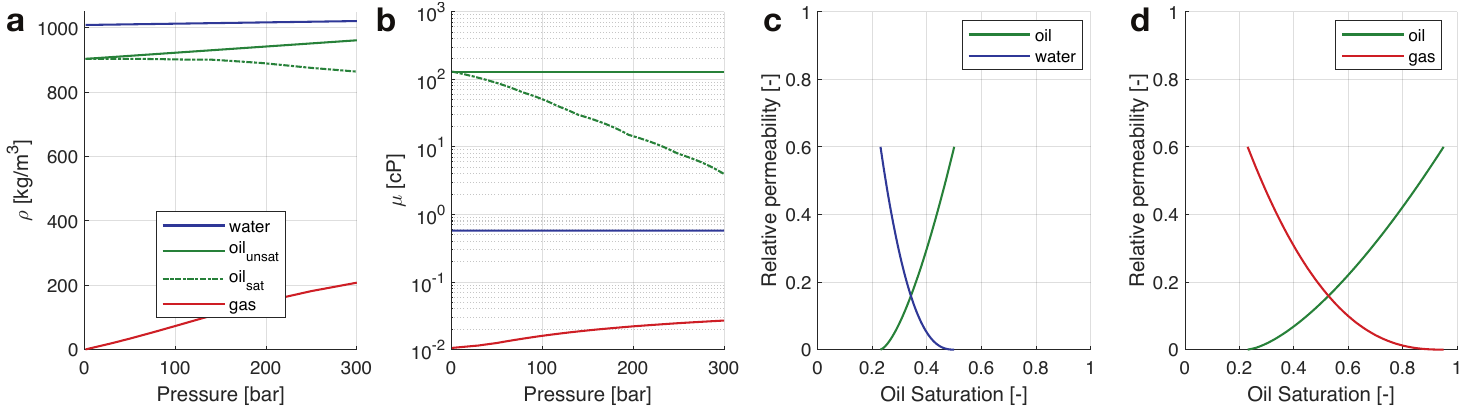}
            \caption{Density (\textbf{a}), dynamic viscosity (\textbf{b}), and two-phase relative permeability curves (\textbf{c}, \textbf{d}) for water, oil, and gas at Wilmington.} 
             \label{fig:fluids}
        \end{figure}

        We initialized the reservoir simulation model with an oil-water contact (OWC) at 1800 m depth, about 1 km below the crest of the Ranger anticline. The oleic phase was assumed to be fully saturated in gas, and no gaseous phase was initially present~\cite{doggr1992,otott1996}. The density and viscosity of the oil and two-phase relative permissibilities were initially based on the SPE9 comparative solution project~\cite{killough1995}, but they were adjusted to better fit the hydrocarbon properties at Wilmington during history matching, and are provided in Fig.~\ref{fig:fluids} (see Table~\ref{tab:zones}).

\begin{figure}[h]    
            \centering
            \includegraphics[width=1.0\textwidth]{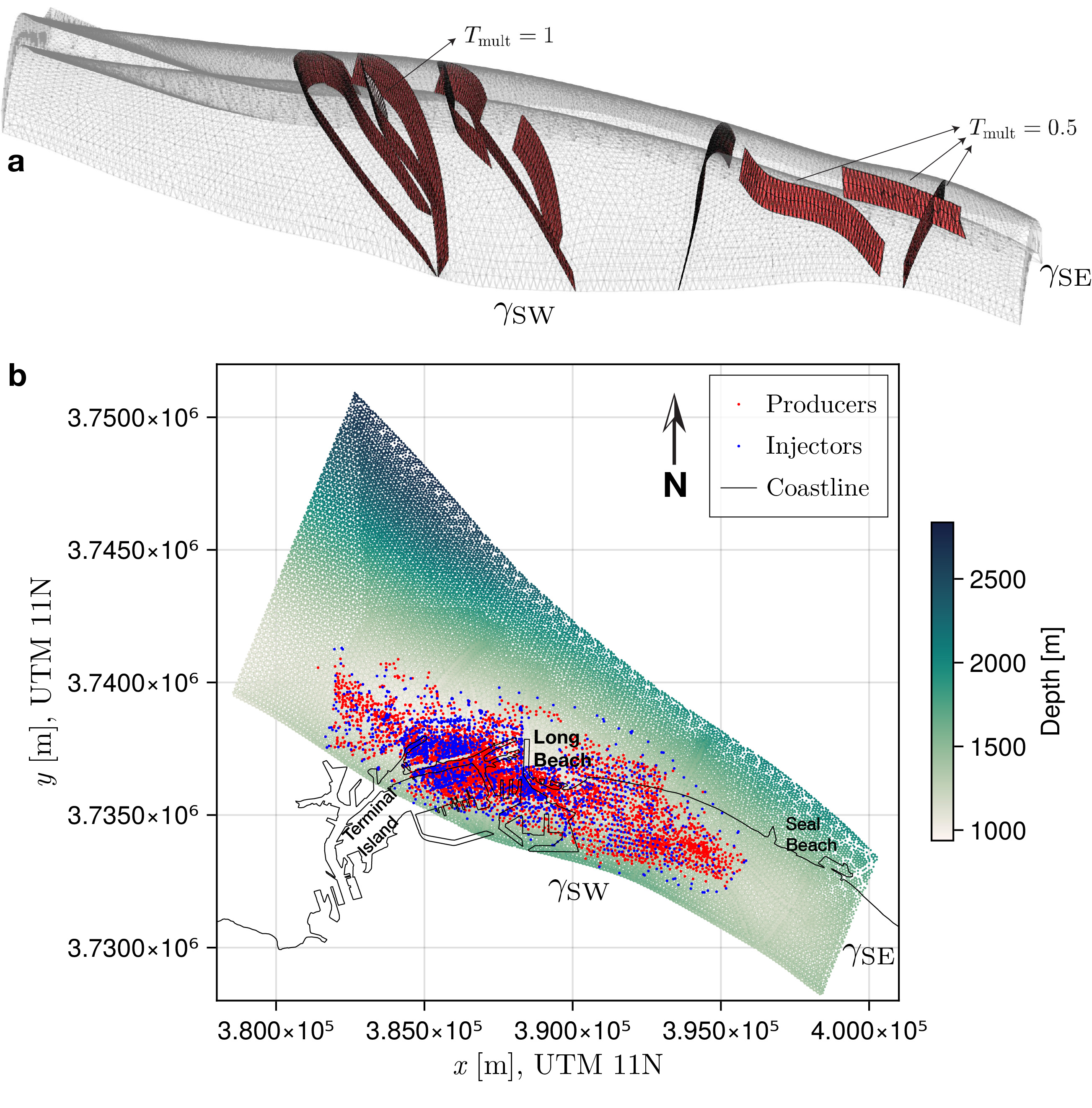}
            \caption{\textbf{a}:~Side view of the reservoir simulation domain showing all tear faults. Transmissibility multipliers for faults with values different than 0, and the south-west and south-east boundaries, where pore volume multipliers were used, are specified. \textbf{b}:~Map view of the reservoir simulation domain, showing cell centroid depths, well locations, and coastline.} 
             \label{fig:faults&wells}
        \end{figure}

        The boundary conditions for the reservoir simulation were set to:
        \begin{equation}
            \begin{aligned}
                & \mathbf{q} \cdot \mathbf{n} = 0,  && \text{on } \Gamma = \bigcup_{i} \gamma_{i}, \\
                & \mathrm{PVM} = 10^4, && \text{on } \gamma_\text{SE},\gamma_\text{SW} \\
            \end{aligned}
        \end{equation}
        i.e., no-flow was specified everywhere, but a pore volume multiplier (PVM) was added to all cells in the south-east and south-west boundaries during history matching to represent formation continuity beyond the domain. Intra-reservoir faults (also referred to as tear or transverse faults) were tested with multiple transmissibilities. Ultimately, the majority were included as sealing (transmissibility multiplier, $T_\text{mult}$, set to 0 on all fault faces), while a small subset were modeled as permeable ($T_\text{mult} = 0.5$ or 1), as shown in Fig.~\ref{fig:faults&wells}a. 
        
        The well data originally downloaded from the California Department of Conservation \footnote{\url{https://www.conservation.ca.gov} \label{note:doc}} contained the latitude, longitude, and monthly rates for 9050 producers and 3073 injectors. However, when multiple wells fell into the same cell, we collapsed them into a single completion, and added their monthly rates. The total number of producers and injectors in our model is 4543 and 1770, respectively (Fig.~\ref{fig:faults&wells}b). 
        
        \subsubsection{Geomechanics model} \label{sec:setup-geomechanics}
\begin{figure}[h]    
            \centering
            \includegraphics[width=1.0\textwidth]{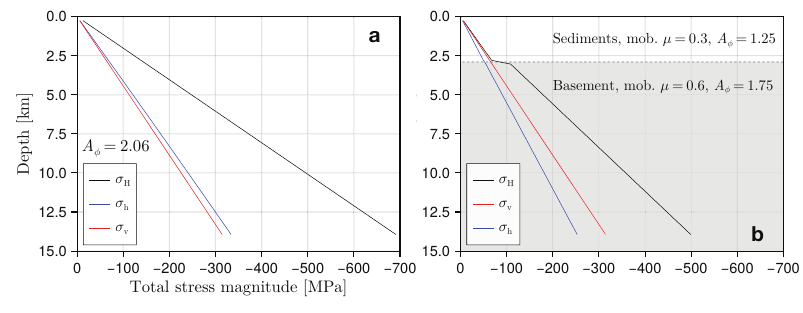}
            \caption{Initial total stress magnitudes as a function of depth. Note that the overburden stress remains the same across the two plots, but its magnitude relative to the other two principal stresses does not. \textbf{a}:~Reverse faulting stress regime, with $A_\phi$ from best fit to data in the SCEC Community Stress Model~\cite{hardebeck2023}. \textbf{b}:~Heterogeneous strike-slip stress regime, where the sedimentary section has lower deviatoric stress (mobilized $\mu$=0.3), and the basement is critically stressed. The depth boundary shown here is representative of the center of our computational domain only, because the thickness of the sedimentary section varies with $x$ and $y$.} 
             \label{fig:stress}
        \end{figure}

        The geomechanics model was initialized by analytical calculation of the background, pre-production stress state: The vertical or overburden stress can be estimated as~\cite{jaeger2007}:
        \begin{equation}
            \sigma_\text{v} = -\int \rho_\mathrm{b}(z) \mathbf{g}(z) \, \mathrm{d}z \approx -\overline{\rho}_\mathrm{b} g z
        \end{equation}
        where $\overline{\rho}_\mathrm{b}$ is the mean bulk density, taken to be 2300 kg/m$^3$ here. According to the Mohr-Coulomb failure criterion, the ratio between the maximum and minimum principal effective stresses at the limit of frictional failure can be estimated as~\cite{jaeger2007}:
        \begin{equation} \label{eq:sratio}
            \frac{\sigma_\text{1}^\prime}{\sigma_\text{3}^\prime} = \frac{\sigma_\text{1} + p_\text{f}}{\sigma_\text{3} + p_\text{f}} = [ (\mu^2 + 1)^{1/2} + \mu]^2 \approx 3.12
        \end{equation}
        where $\mu$ is the friction coefficient, taken to be 0.6~\cite{byerlee1978}. Using Eq.~\ref{eq:sratio} with a given value for the friction coefficient and known pore fluid pressure, $\sigma_1$ or $\sigma_3$ can be approximated once $\sigma_\text{v}$ is known. $\sigma_2$ can also be estimated by selecting a value of $A_\phi$, a parameter proposed by~\citeA{simpson1997} to express the relative size of the three principal stresses:
        \begin{equation} \label{eq:aphi}
            \begin{aligned}
            A_\phi & = (n + \frac{1}{2}) + (-1)^n(\psi-\frac{1}{2}) \\[2mm]
            \psi & = \frac{\sigma_2 - \sigma_3}{\sigma_1 - \sigma_2}
            \end{aligned}
        \end{equation}
        where $n=0$ for normal faulting, $n=1$ for strike-slip faulting, and $n=2$ for reverse faulting. Values of $A_\phi$ between 0 and 1 indicate normal faulting conditions; between 1 and 2 strike-slip; and between 2 and 3 reverse faulting. Hence, the exact form of the equations to estimate the value of each principal stress magnitude depend on the stress regime, and are given by~\citeA[ch. 7; note their positive compression sign convention]{lundsnee2020}. Fig.~\ref{fig:stress}a,b shows the principal stress magnitudes used to initialize the models discussed in \S~\ref{sec:results}, where $\sigma_\text{H}$ and $\sigma_\text{h}$ refer to the maximum and minimum horizontal stresses.

        The boundary conditions for the geomechanics model were set to:
        \begin{equation}
            \begin{aligned}
                & \overline{\mathbf{u}} = \mathbf{0}, && \text{on } \gamma_{\mathrm{bot}} \\
                & u_x = 0, \;\; u_y = 0, && \text{on } \gamma_{\mathrm{NW}} \cup \gamma_{\mathrm{NE}} \cup \gamma_{\mathrm{SE}} \cup \gamma_{\mathrm{SW}} \\
                & \boldsymbol{\sigma}\cdot\mathbf{n} = \mathbf{0}, && \text{on } \gamma_{\mathrm{top}}
            \end{aligned}
        \end{equation}
        i.e., the bottom boundary is fixed to 0 displacement, the side boundaries have both horizontal degrees of freedom fixed, and the top surface is traction-free. Note that all side and bottom boundaries in the geomechanics model are located several km away from the production interval, and therefore they do not influence the computed strains. 
        
\begin{figure}[h!]    
            \centering     \includegraphics[width=0.7\textwidth]{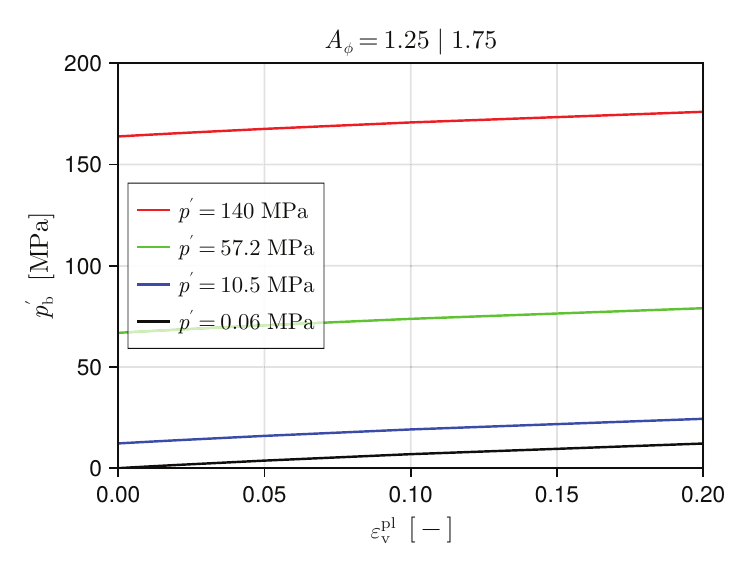}
            \caption{Hardening curves obtained with Eq.~\ref{eq:hardening} and the parameter values listed in Table~\ref{tab:constitutive-parameters} (strike-slip model). $\varepsilon_\text{v}^{\text{pl}}$ is the volumetric plastic strain. Values of $p^\prime$ cover the range encountered in the simulations.} 
             \label{fig:hardening}
        \end{figure}

        The constitutive parameters are provided in Table~\ref{tab:constitutive-parameters}; the hardening curves for $p^\prime_b$ (Eq.~\ref{eq:pa}, Fig.~\ref{fig:dpcm}) were computed as:
        \begin{equation} \label{eq:hardening}
            p^\prime_b \; (p^\prime, \varepsilon_\text{v}^{\text{pl}}) = a p^\prime + \frac{b}{3D} \ln \left( \frac{\varepsilon_\text{v}^{\text{pl}}}{W} + 1 \right)
        \end{equation}
        where $D$ and $W$ are material parameters based on the work of~\citeA{vesic1968,kosloff1980b}, and $a$ and $b$ are fitting parameters to be adjusted based on the initial state of stress and cap shape, and on the amount of volumetric plastic deformation obtained, respectively (Table~\ref{tab:constitutive-parameters}). Representative curves for different values of $p^\prime$ are provided in Fig.~\ref{fig:hardening}.
            
        \begin{table}[htbp]
            \centering
            \caption{Constitutive model parameters for the history-matched model presented in \S~\ref{sec:results}. The sedimentary section includes the Ranger and Pico Zones, as well as Pico's overburden and sediments below the Ranger Zone. The basement and ``other" sedimentary units (above the basement, to the sides of the producing units) were modeled as purely elastic.}
            \label{tab:constitutive-parameters}
            \begin{tabular}{lccc}
                \hline
                & {$A_\phi=1.25 \;|\; 1.75$} & & \\
                \hline 
                 Parameter & Sedimentary Section & Basement & Other \\
                \hline
                $E$ [GPa] & 0.6 & 20 & 0.1 \\
                $\nu$ & 0.25 & 0.25 & 0.25 \\
                $d$ [kPa] & 1 &  & \\
                $\beta$ [$^\circ$] & 55 & & \\
                $R$ & 0.8 & & \\
                $D$ [GPa]$^{-1}$ & 7.14 & & \\
                $W$ & 0.27 & & \\
                $a$ & 1.17 & & \\
                $b$ & 0.47 & & \\
                \hline
            \end{tabular}
        \end{table}

        We note that faults are included as explicit geologic surfaces in the mesh. Their   hydraulic properties influence fluid flow, and, as a result, geomechanical response. We do not, however, solve a contact problem on the fault surfaces in this model, meaning that faults are in ``stick" mode or static throughout the simulation, and are included as nodesets in Abaqus. We then use the stress tensors reported by Abaqus along the fault nodesets to quantify fault stability (see \S~\ref{sec:fault_stability}). 

\section{Results and Discussion} \label{sec:results}
    \subsection{Simulation of hydrocarbon production and pore pressure response} \label{sec:results_reservoir}
\begin{figure}[h]    
            \centering
            \includegraphics[width=1.0\textwidth]{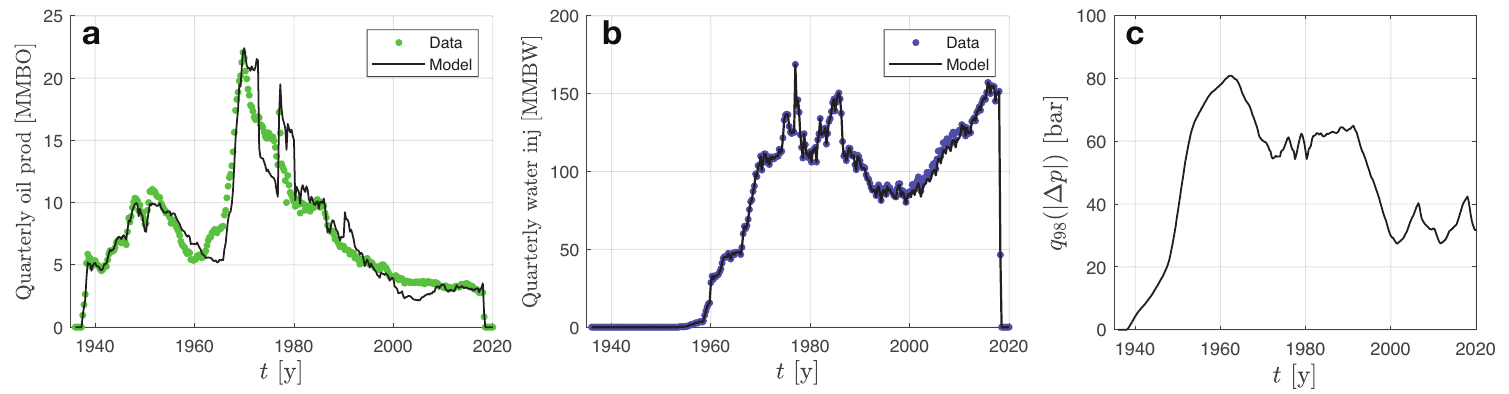}
            \caption{\textbf{a}: Quarterly oil production in millions of barrels of oil (MMBO). \textbf{b}: Quarterly water injection in millions of barrels of water (MMBW). \textbf{c}: 98th percentile of the absolute value of pore pressure change between a given time and the pre-production pressures. The percentile was computed using cells above the oil-water contact only.} 
             \label{fig:fluid_prod}
        \end{figure}

        Our simulation of hydrocarbon production at Wilmington covers 85 y between 1936 and 2020. Given that the well completion depths are not public (\S~\ref{sec:setup}), all of the producers and injectors were placed in our reservoir model domain, which is $\approx 200$ m thick. This is representative of the combined thickness of the Ranger and Upper Terminal Zones, which are the most productive (\S~\ref{sec:field}). We specified liquid well rates for both producers and injectors, based on the dataset downloaded from the California Department of Conservation (see footnote~\ref{note:doc}). Fig.~\ref{fig:fluid_prod}a,b shows the total oil production and water injection rates for each quarter, comparing the specified well rates (data) and the actual well rates after model convergence (model). Small differences exist, particularly in oil production, because some wells switch from rate-controlled to bottom-hole-pressure (BHP)-controlled once the BHP reaches very low ($\rightarrow0$) or very high (100 bar above hydrostatic) pressures. The cumulative oil production from our model is $\approx 2.49$ billion barrels, consistent with the value of 2.5 billion barrels widely reported~\cite<e.g.>{kwong2019}.

\begin{figure}[h]    
            \centering
            \includegraphics[width=1.0\textwidth]{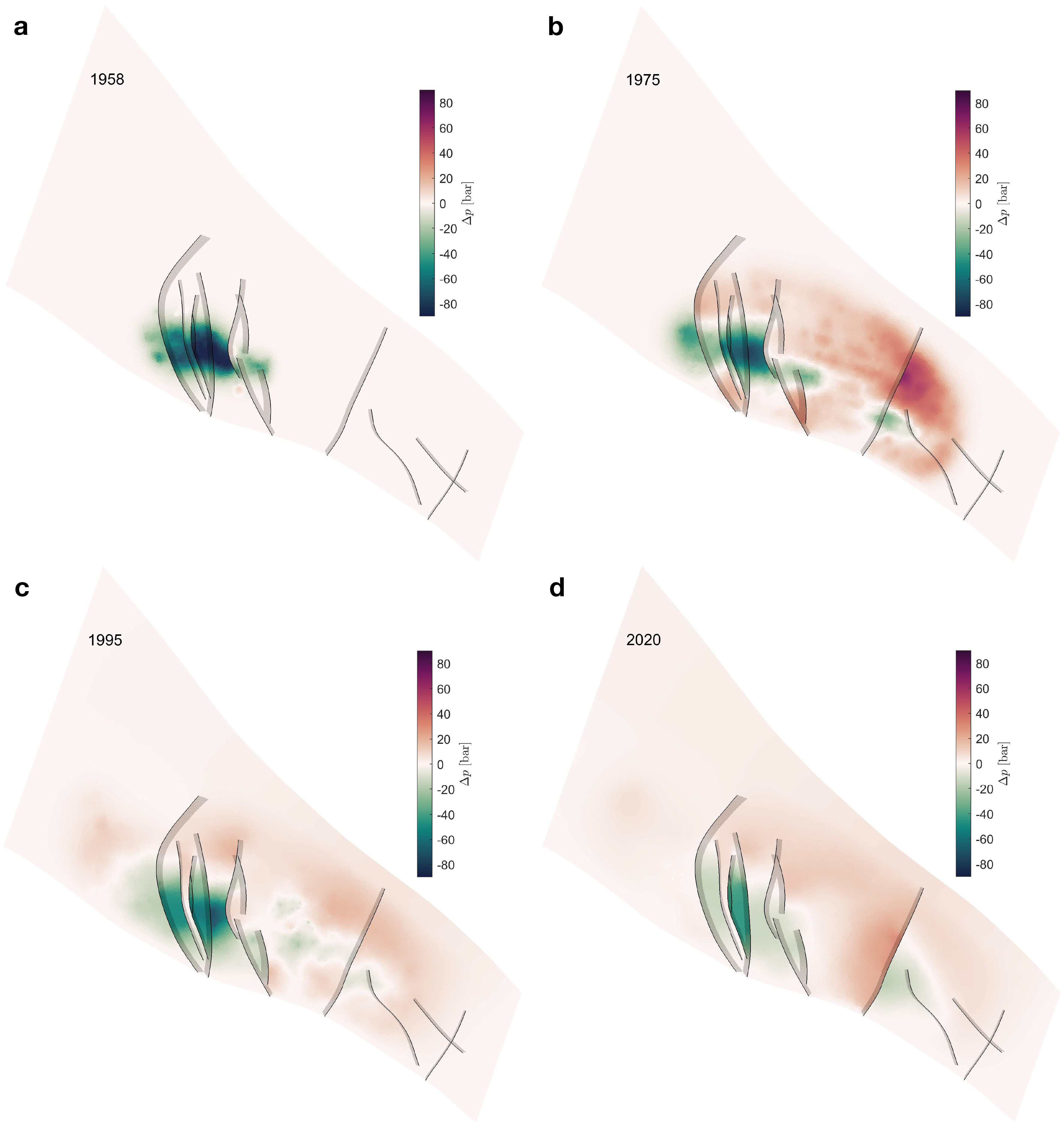}
            \caption{Pressure changes (with respect to pre-production pressures) in the Ranger Fm in (\textbf{a}) 1958, before water injection started; \textbf{b}: 1975, after re-pressurization; \textbf{c}: 1995; and \textbf{d}: 2020.} 
             \label{fig:dp_maps}
        \end{figure}

        Changes in fluid pressure ($\Delta p$), with respect to pre-production values, are illustrated in Fig.~\ref{fig:fluid_prod}c and Fig.~\ref{fig:dp_maps}. Absolute values of pressure change were highest during the first part of field development (1940-1960). During this time, the pore pressure was depleted in the center of the field, where subsidence occurred; our model predicts local pressure changes up to 80-90 bar. The ellipse of pore pressure reduction $>10$ bar in Fig.~\ref{fig:dp_maps}a has dimensions of $\approx45$ km$^2$, which is similar to the reported 50-55 km$^2$ for the subsidence bowl~\cite{mayuga1970,pierce1970}. The spatial extent of pore pressure depletion is relatively limited due to the sealing faults within the reservoir. Between 1960 and 1980, enhanced oil recovery via water-flooding and expansion to the offshore part of the field led to the maximum quarterly oil production (Fig.~\ref{fig:fluid_prod}a). We observe pressurization in several areas around the two main production lobes, although the absolute pressure differences (with respect to pre-production levels) remain lower than during the first part of reservoir operations. Pore pressure changes appear to be further limited in the 1980-2020 period, as can be observed in Fig.~\ref{fig:dp_maps}c, d.

        It is difficult to further calibrate the reservoir simulation model without well pressure data, which are not publicly accessible via the California Department of Conservation. Field pressure data available online include average reservoir pressures before and after the initial production period (1930s to 1960s). For example, the~\citeA{dog1964} report pre-production reservoir pressure as well as average reservoir pressure in 1958-1960 for the Ranger and Upper Terminal Zones, showing pressure depletion in excess of 80 bar in some fault blocks. Average reservoir pressures, at multiple times in the period 1936-1960, were reported by~\citeA{yang1998} for the Tar Zone, fault block II-A, which show a decrease of more than 60 bar in the period 1936-1960. These data are consistent with our model outputs, which predicts maximum local pore pressure decreases of 80-90 bar, bringing the fluid pressure, in certain cells, to values that are barely above atmospheric.
        
    \subsection{Background stress regime and ground deformation} \label{sec:results_mech}
        The stress state is typically uncertain in the shallow crust~\cite<$< 5$~km depth; e.g.,>{lundstern2020,delorey2021}. In the Los Angeles Basin, in particular, the mixed-faulting style introduces significant uncertainty despite relatively consistent orientations of $\sigma_\text{H}$ and $\sigma_\text{h}$ (see \S~{\ref{sec:stress}}). Given that the initial stress state directly influences deformation in elastoplastic formations (\S~\ref{sec:poromech}), we explore a range of initial stress conditions compatible with the data summarized in \S~\ref{sec:stress}. The initial stress conditions of the two models discussed in this section are shown in Fig.~\ref{fig:stress}.
        
\begin{figure}[h!]    
            \centering
            \includegraphics[width=1.0\textwidth]{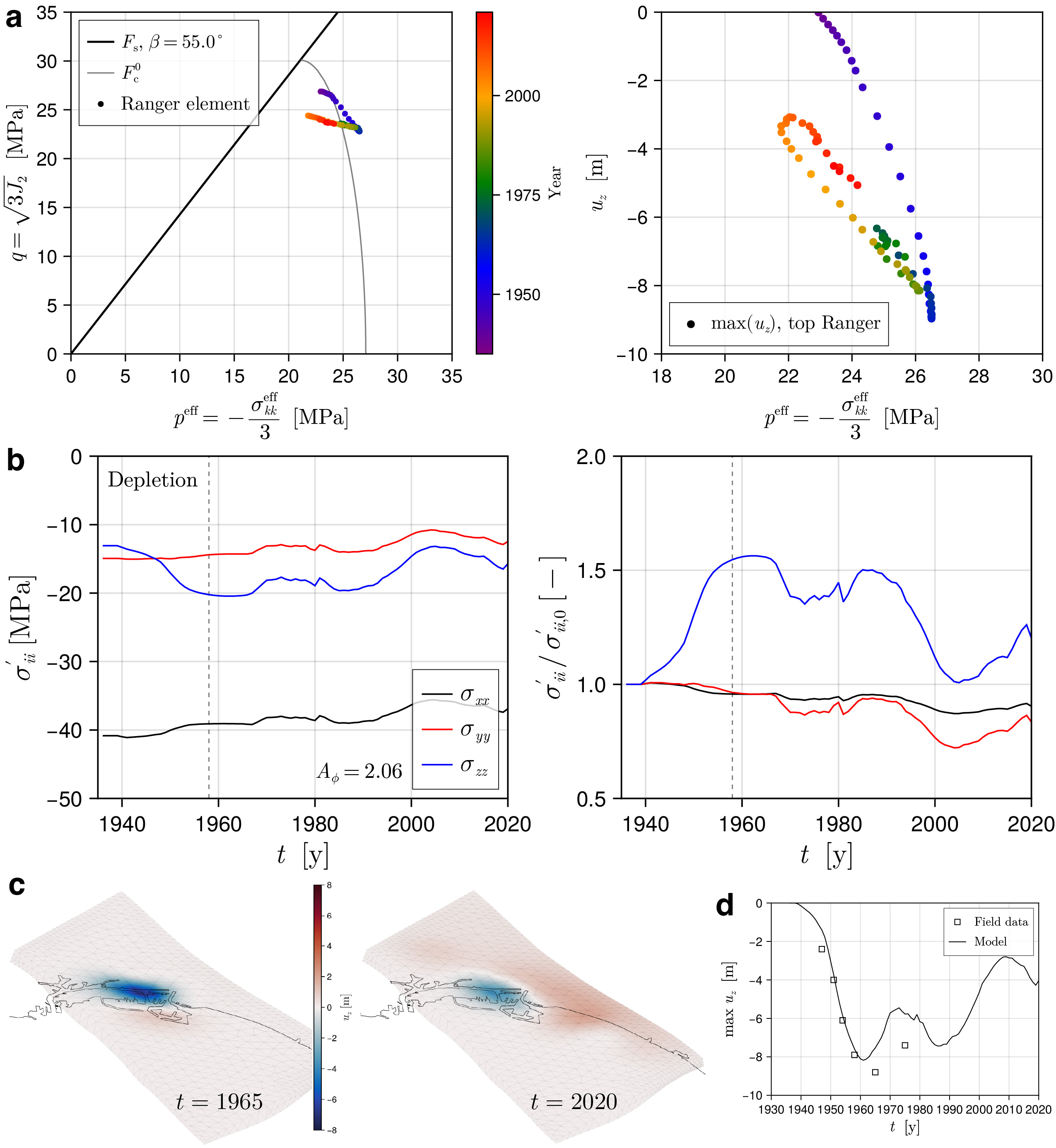}
            \caption{Summary of geomechanical results for the reverse-faulting model ($A_\phi = 2.06$; see \S~\ref{sec:setup-geomechanics}  and Fig.~\ref{fig:stress}a). \textbf{a}: Stress path (left) and vertical displacement computed at the center of the subsidence bowl. \textbf{b}: Temporal evolution of the horizontal and vertical effective stress components aligned with the grid axes (left) and ratios with their initial values (right). \textbf{c}: Surface vertical displacements, with respect to pre-production levels, in 1965 and 2020. \textbf{d}: Temporal evolution of displacement at the node where maximum subsidence was computed, compared with maximum levels registered in the field.} 
             \label{fig:reverse}
        \end{figure}

        The geomechanics results for the critically-stressed model initialized with a reverse faulting regime ($A_\phi = 2.06$; see Fig.~\ref{fig:stress}a) are summarized in Fig.~\ref{fig:reverse}. In Fig.~\ref{fig:reverse}a,b it can be seen that, during the initial stage of reservoir production, the deviatoric stresses decrease in the Ranger zone. This is primarily due to the evolution of $\sigma^\prime_{zz}$, which registers the largest change, and means that very small adjustments of the cap are necessary (Fig.~\ref{fig:reverse}a). As a result, matching the modeled subsidence to the field data requires a nearly vertical cap ($R<0.1-0.2$), virtually no hardening ($b<0.1$), and a very soft sedimentary sequence ($E\lesssim0.3$ GPa). These values, however, lead to overestimation of the rebound of the ground surface associated with water injection (Fig.~\ref{fig:reverse}c,d). 

\begin{figure}[h!]    
            \centering
            \includegraphics[width=1.0\textwidth]{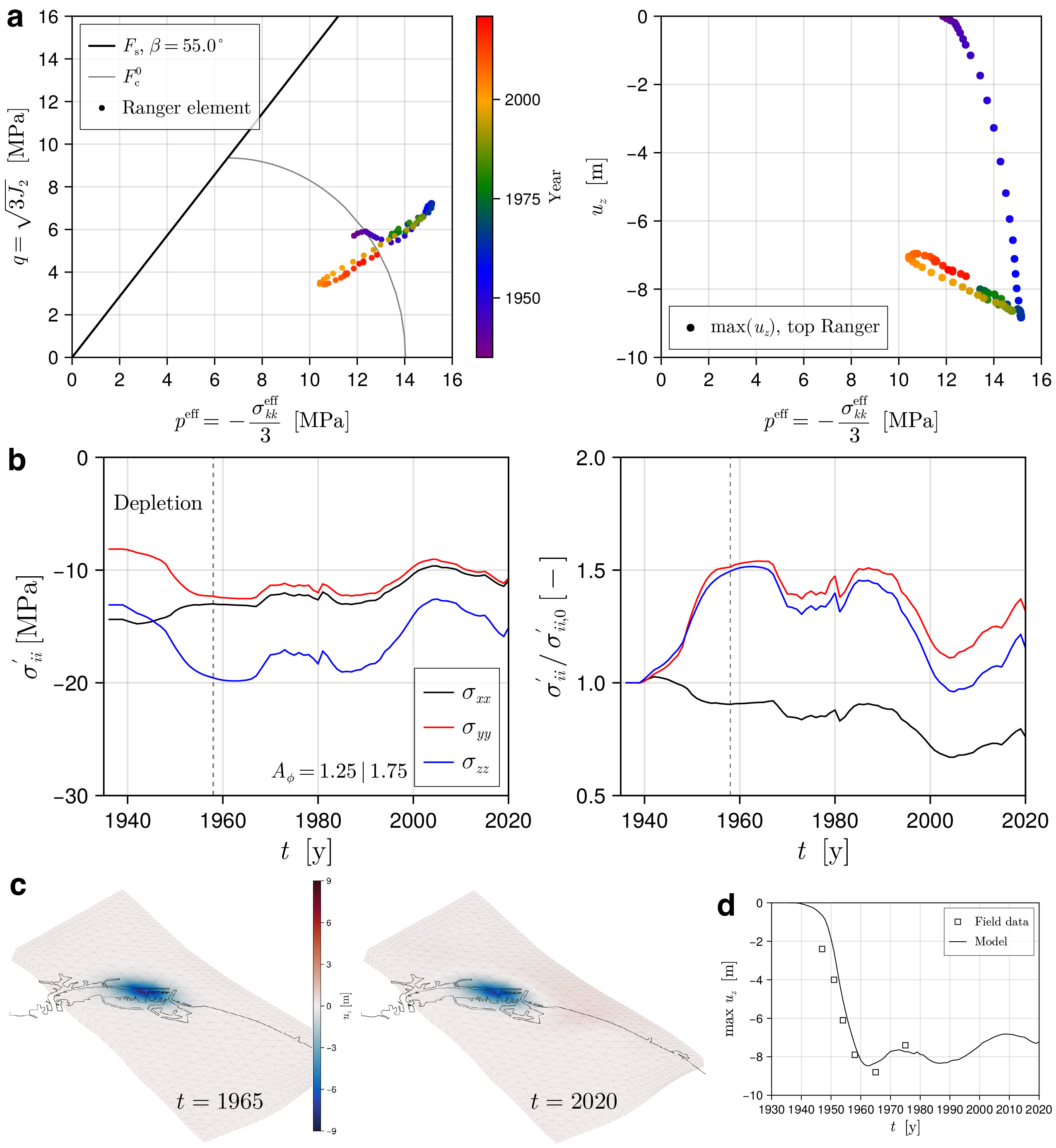}
            \caption{Summary of geomechanical results for the strike-slip model (see \S~\ref{sec:setup-geomechanics} and Fig.~\ref{fig:stress}b). \textbf{a}: Stress path (left) and vertical displacement computed at the center of the subsidence bowl. \textbf{b}: Temporal evolution of the horizontal and vertical effective stress components aligned with the grid axes (left) and ratios with their initial values (right). \textbf{c}: Surface vertical displacements, with respect to pre-production levels, in 1965 and 2020. \textbf{d}: Temporal evolution of displacement at the node where maximum subsidence was computed, compared with maximum levels registered in the field.} 
             \label{fig:ss}
        \end{figure}

        In contrast, Fig.~\ref{fig:ss} shows the results of the model initialized with a low deviatoric stress in a strike-slip faulting regime. The combination of low mobilized friction  and $A_\phi = 1.25$ in the sedimentary section during initialization results in the maximum horizontal stress and vertical stress having relatively similar magnitudes (Fig.~\ref{fig:stress}b). Note that, in contrast to the basement, which is typically critically stressed, sedimentary basins are not~\cite{juanes2012,vilarrasa2015}. During the first stage of production, the vertical stress becomes the maximum stress component, leading to an increase in the deviatoric stress. This type of trajectory requires much larger adjustments of the cap, which means that matching the modeled subsidence to the field data can be accomplished with more reasonable parameter values (see Table~\ref{tab:constitutive-parameters} and discussion below), and the rebound triggered by increased water injection agrees with field observations (Fig.~\ref{fig:ss}c,d). Therefore, the first-order control on subsidence and rebound is the stress state of the sedimentary formations where reservoir operations took place. Note that, in the model in Fig.~\ref{fig:ss}, the basement remains critically-stressed (Fig.~\ref{fig:stress}b); even if the absolute stress levels are lower than in a reverse-faulting regime, we expect that similar surface displacements can be obtained if the basement were initialized with a critically-stressed, reverse faulting regime.  

        As noted in \S~\ref{sec:poromech}, $R$ controls the eccentricity of the cap in $p^\prime$--$q$ space (i.e., the extent to which the onset of plastic compaction depends on deviatoric stress). In the limit $R\rightarrow 0$, the cap approaches a vertical line in $p^\prime$--$q$ space, implying that the mean effective stress required to activate plastic compaction is independent of $q$. This is inconsistent with critical-state soil mechanics, in which yield and volumetric plastic response depend on the stress ratio and stress path, rather than $p^\prime$ alone~\cite{sch68}. Consistent with this, modern constitutive models for sand define yielding and plastic flow in terms of the stress ratio $\eta=q/p^\prime$, so that yielding and dilatancy/contraction depend on $\eta$~\cite<e.g.,>{jefferies1993}. Direct evidence that deviatoric loading contributes to irreversible compaction in sands is provided by \citeA{vesic1968}, who performed ``octahedral shear'' tests in which $q$ was increased while the mean normal stress was held approximately constant. These tests show volume decrease during shear and enhanced grain crushing compared with isotropic compression, demonstrating that plastic compaction cannot be treated as a purely hydrostatic phenomenon. Therefore, although we infer $R$ for the Wilmington sedimentary cover by history matching ground deformations, very small cap eccentricities are not compatible with the well-established observation that deviatoric stress meaningfully contributes to compaction in granular materials.

        The cap hardening (Eq.~\ref{eq:hardening}) models how the cap position on the hydrostatic stress axis, $p^\prime_b$, evolves as a function of the plastic volumetric strain $\varepsilon_\text{v}^\text{pl}$. When yielding on the cap, plastic compaction leads to hardening (i.e., further compaction requires larger stresses). For example, \citeA{vesic1968} report isotropic compression curves for sand at mean stresses up to $\sim 630~\mathrm{kg/cm^2}\approx 620$~bar, showing that attaining large volumetric compaction ($10$--$20\%$) requires stress increases that are progressively larger. This is consistent with compression models for sands under hydrostatic and one-dimensional compression~\cite<e.g.,>{pestana1995}, which show high nonlinearity. In particular, both of these studies suggest that a multi-fold increase in effective mean stress is necessary to achieve significant levels of plastic compaction, especially with grain crushing. In the reverse faulting model, we require values of $b$ in Eq.~\ref{eq:hardening} that are very small, leading to hardening of only a few percent over the range $\varepsilon_\text{v}^\text{pl}=0$--20\%, and therefore inconsistent with laboratory measurements in sands.

        Moreover, in sands, laboratory tests indicate that elastic moduli  are on the order of several hundred MPa and up to 1~GPa at effective confinement levels relevant for Wilmington production depths~\cite<e.g.>{vesic1968}. Unloading data after compaction at elevated stresses show small elastic strain, consistent with a stiff elastic response~\cite<e.g.>{pestana1995,zheng2021}. Therefore, values of $E\sim 0.2$~GPa, as in Fig.~\ref{fig:reverse}, are likely not representative of Wilmington materials, consistent with overestimation of rebound in the simulations with a reverse-faulting stress regime. Taken together, the parameter values for the elastic and plastic response that we require to match surface deformation to observed values are reasonable in the model initialized with low deviatoric stress in the sedimentary section, but much less so in the reverse-faulting model, particularly if the sedimentary sequence is also critically stressed.
        
    \subsection{Fault stability analysis} \label{sec:fault_stability}
        Fluid extraction and injection leads to changes in (i) the effective normal stress, $\sigma^\prime_\text{n}$ (via pore pressure changes as well as poromechanical deformation) and (ii) the shear stress, $\tau$ (via poromechanical deformation) acting on preexisting fault surfaces. The combined effect of these changes on fault stability can be quantified by means of {\em changes} in the Coulomb Failure Function (CFF) defined as~\cite{reasenberg1992,king1994,salo2017,hager2021,silva2021,silva2023}:
        \begin{equation}
            \begin{aligned}
                \text{CFF} &= \vert \tau \vert + \mu\sigma^\prime_\text{n}, \\
                \text{DCFF}(t) &= \text{CFF}(t) - \text{CFF}(t_0)\\
            \end{aligned}
            \label{eq:dcff}
        \end{equation}
        where $t_0$ refers to the initial condition (before hydrocarbon production), and DCFF is the delta or change in CFF. Although alternatives to the DCFF framework exist~\cite<e.g.,>{meade2017}, we use it here due to its popularity and straightforward physical interpretation. In the sign convention adopted throughout this paper, which is that tension is positive, $\text{DCFF} > 0$ indicates that the corresponding fault is destabilized (i.e., brought closer to failure). To calculate DCFF values, we used $\mu=0.6$~\cite{byerlee1978}.

\begin{figure}[h]    
            \centering
            \includegraphics[width=1.0\textwidth]{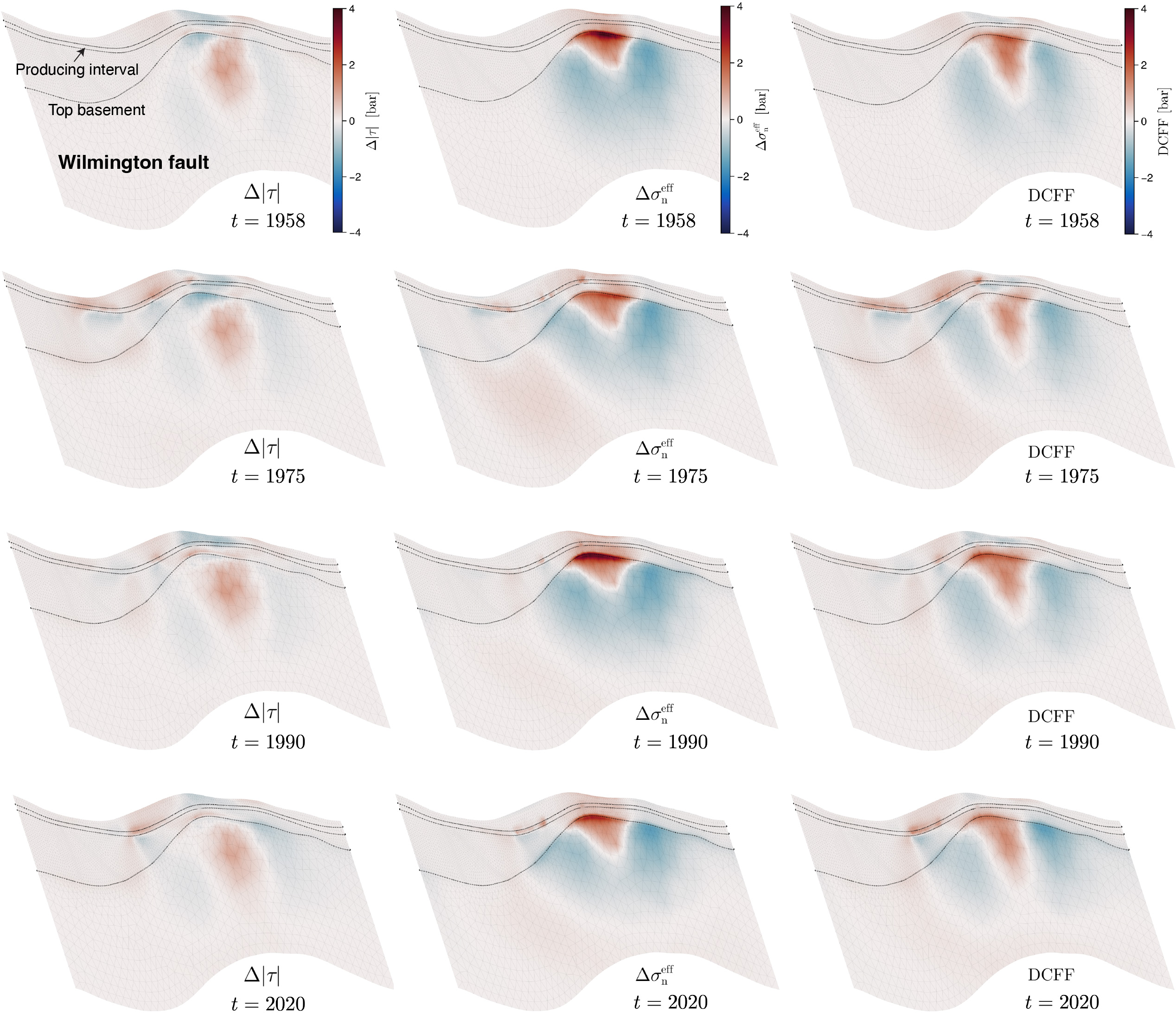}
            \caption{Stress changes (with respect to 1936) mapped on the Wilmington Fault. Left column: shear stress change; Center column: Effective normal stress change; Right column: DCFF. Top row: 1958; Second row: 1975; Third row: 1990; Bottom row: 2020.} 
             \label{fig:wf}
        \end{figure}

        We now focus on the evolution of fault stress changes in the strike-slip background stress model, initialized with low deviatoric stress in the sedimentary section (Fig.~\ref{fig:ss}; \S~\ref{sec:results_mech}). The evolution of stress changes on the Wilmington Fault is presented in Fig.~\ref{fig:wf} (refer to Fig.~\ref{fig:GeologicalModelFromGocad} for location). It can be seen that pore pressure depletion during the first part of production (1936-1958) leads to destabilization in the central part of the fault, in the upper part of the basement. The main reason for this destabilization is the reduction in effective normal stress due to reservoir compaction. The repressurization program (1958-1975) led to localized destabilization in the Wilmington Fault inside the production depth range, in areas closer to the injection wells where the pore pressure increased. Overall, we find that the stress changes incurred during the first part of production (1936-1958) dominate fault stability throughout the present. We note, however, that the maximum magnitude of DCFF on the Wilmington Fault is $<~4$ bar, which is a small perturbation at production depths or deeper.

\begin{figure}[h]    
            \centering
            \includegraphics[width=1.0\textwidth]{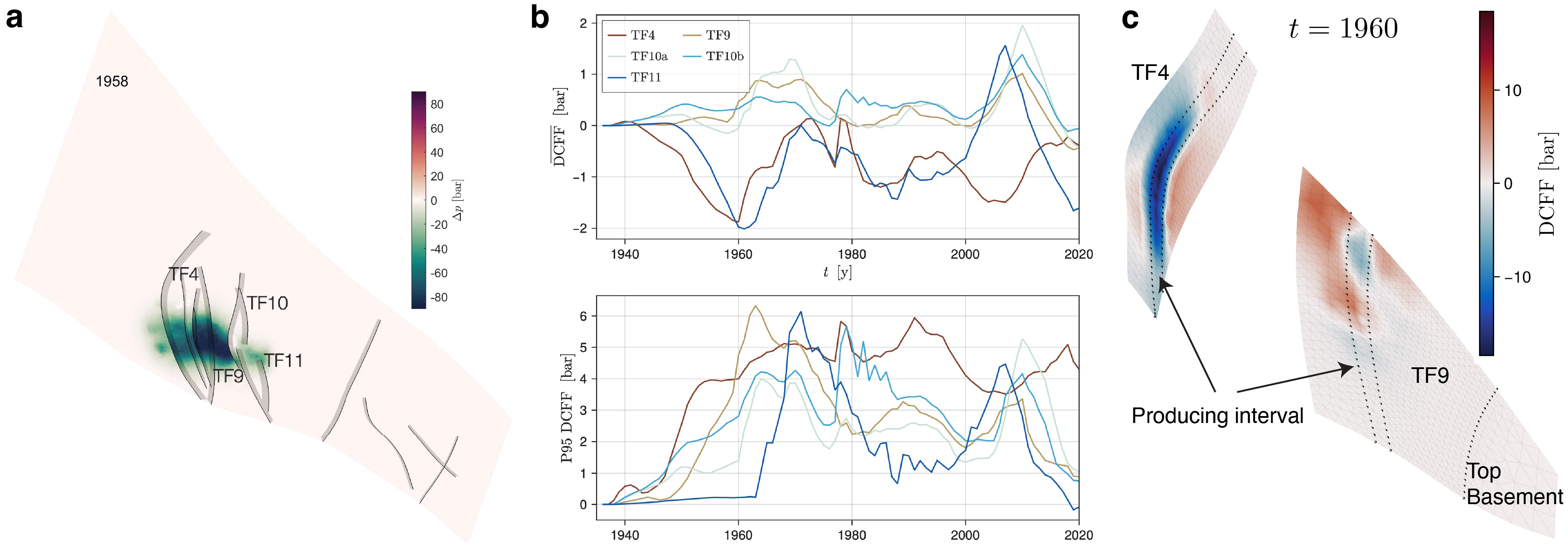}
            \caption{\textbf{a} Location of Tear Faults explored in the reservoir. \textbf{b}: Evolution of average (top) and 95th percentile (bottom) DCFF on TF4, TF9, TF10, and TF11. \textbf{c}: Snapshot of DCFF on TF4 (Cerritos) and TF9 (Allied) in 1960. The maximum DCFF on TF9 is $\approx8.6$ bar.} 
             \label{fig:tf}
        \end{figure}

        Stress changes on the reservoir tear faults are summarized in Fig.~\ref{fig:tf}b, where we show the evolution of mean and 95th percentile of DCFF on four faults in the pore pressure depleted area. On average, the DCFF does not change much ($\vert \overline{\text{DCFF}} \vert \leq 2$ bar). However, there is local destabilization on portions of most faults, with P95 values above 6 bar in TF4, TF9, and TF11. This is expected in  hydrocarbon reservoirs in normal faulting regime with $b = 1$ \cite{segall1989,segall1998}, consistent with the post-depletion stress state of our model initialized with low deviatoric stress (see Fig.~\ref{fig:ss}). Fig.~\ref{fig:tf}c shows a map of DCFF in 1960 on the Cerritos (TF4) and Allied (TF9) faults~\cite{doggr1992}. In the Allied Fault, we obtain maximum values of around 10 bars above the producing interval.

\begin{figure}[h!]    
            \centering
            \includegraphics[width=0.89\textwidth]{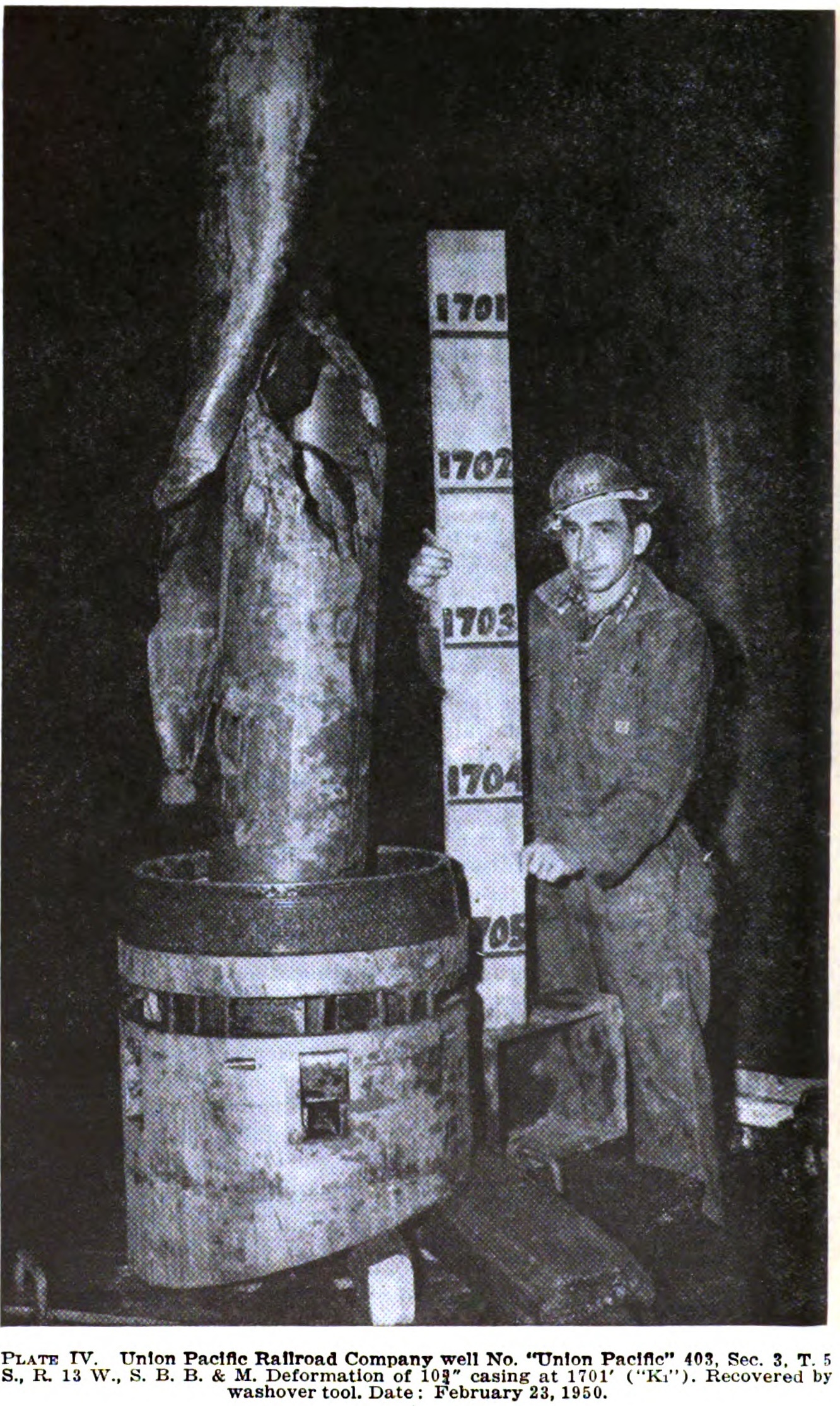}
            \caption{Reprint of the striking picture from \citeA{frame1952}, showing well damage at $\approx$ 500 m depth.} 
             \label{fig:sheared_well}
        \end{figure}

        \citeA{frame1952,kovach1974} documented damage due to shallow earthquakes at Wilmington between 1947 and 1961, predominantly at a depth of $\approx$500~m. As discussed in \S~\ref{sec:stress}, these earthquakes were documented to occur on horizontal bedding planes, reportedly due to shear stress increase on the flanks of the subsidence bowl. \citeA{frame1952}, however, also reported that detailed field investigation found that 117 wells had been damaged at depths ranging from $\approx$ 200 m to $\approx$ 1200 m, and included photographic evidence (Fig.~\ref{fig:sheared_well}). Although the casing may have been damaged further by the recovery pull, it is interesting to note that the offset plane in Fig.~\ref{fig:sheared_well} is steeply dipping, rather than horizontal.

\begin{figure}[h!]    
            \centering
            \includegraphics[width=1.0\textwidth]{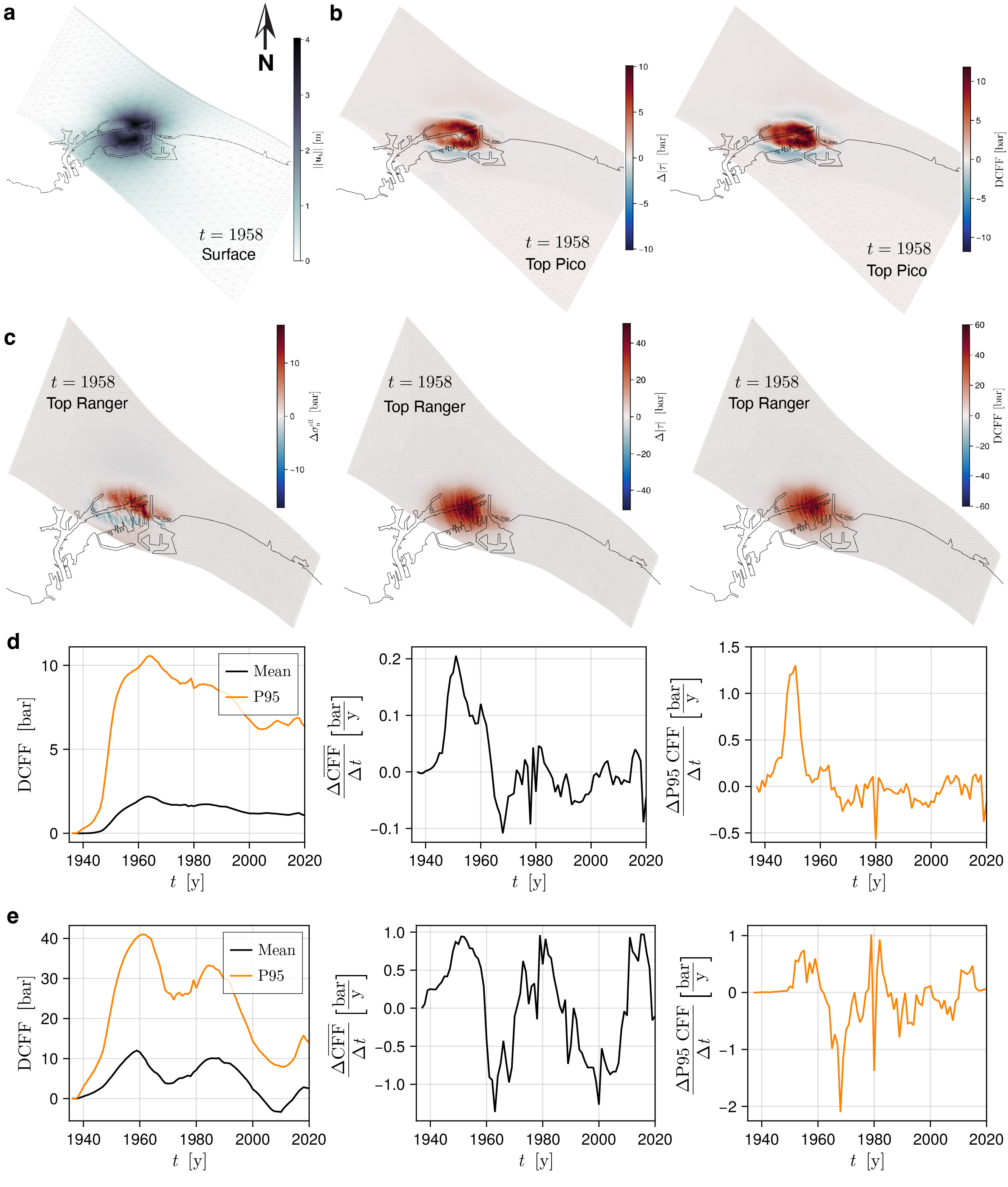}
            \caption{\textbf{a} Magnitude of computed surface horizontal displacement ($|| \boldsymbol{u}_\text{h} ||$). A maximum of 4.03 m is in close agreement with ground observations, documenting $3.5-4$ m (\S~\ref{sec:ground_def}). \textbf{b}: Change in shear stress magnitude (left) and DCFF (right) on the Top Pico surface. \textbf{c}: Change in effective normal stress (left) shear stress magnitude (center) and DCFF (right) on the Top Ranger surface. \textbf{d} Time series of mean and P95 DCFF, and the yearly rate of change of mean and P95 CFF, on the Top Pico surface. To compute these values, we only used locations where surface $|| \boldsymbol{u}_\text{h}||\geq1$ m. \textbf{e}: Same as \textbf{d}, but now for the Top Ranger surface.} 
             \label{fig:pico} 
        \end{figure}

        To investigate this, we use our model to calculate stress changes on two sub-horizontal surfaces at similar depths (in addition to the reservoir faults previously described). To represent the proposed bedding plane faults, we compute the shear stress change and DCFF on the Top Pico and Top Ranger surfaces  (Fig.~\ref{fig:pico}a,b,c). We find that (i) the DCFF is primarily driven by the shear stress change; (ii) the mean DCFF and P95 reach values above 2 and 10 bar (Top Pico), and above 10 and 40 bar (Top Ranger), respectively; (iii) the maximum DCFF values occur towards the end of the pore pressure depletion stage, around 1960; and (iv) the highest rate of CFF change occurs between 1945 and 1960 in the Top Pico surface, but not in the Top Ranger surface (Fig.~\ref{fig:pico}d). Accordingly, our results suggest that:
        \begin{itemize} 
            \item Reservoir operations primarily led to destabilization on sub-horizontal surfaces in the subsidence bowl (Fig.~\ref{fig:pico}), supporting interpretations of seismic fault slip on bedding planes that sheared wells~\cite{frame1952,kovach1974}.
            \item Shallow portions of the reservoir faults were also destabilized (Fig.~\ref{fig:tf}), and therefore could have hosted some of the slip events that led to well damage, consistent with photographic evidence (Fig.~\ref{fig:sheared_well}).
            \item Maximum DCFF was achieved when fault slip and earthquakes associated with subsidence were observed (Fig~\ref{fig:pico}d,e), in agreement with previous interpretations of triggered seismicity~\cite{hager2021}.
        \end{itemize}

    \section{Conclusions}
        
        In this study, we develop the first geologically-realistic model that reproduces Wilmington subsidence and uplift over the full production history (1936–2020). Our main findings are:
        \begin{itemize}
            \item The initial stress state, typically uncertain in the shallow crust ($<5$ km depth), controls reservoir stress paths. The results from our coupled flow-geomechanics poroelastoplastic model indicate that an initially critically-stressed reservoir in a reverse-faulting regime is incompatible with the measurements of ground surface displacements; instead, a strike-slip initialization (lower stress levels) with low mobilized friction in the sedimentary cover provides a better match to observed ground deformation. This implies that the state of stress changes substantially with depth at Wilimington, where the sedimentary section was in a strike-slip regime with low deviatoric stress (or more extensional regime) prior to production, whereas deep basement rocks exhibit a reverse faulting regime based on tectonic earthquakes.
            \item Fault stability analysis indicates localized positive DCFF on intra-reservoir tear faults; however, shear stress and DCFF magnitudes are smaller than those obtained on shallow sub-horizontal surfaces. This suggests that bedding plane faults, in addition to steeply dipping reservoir faults, were destabilized and thus may have caused the observed shearing of wells and recorded earthquakes at Wilmington between 1947 and 1961.
        \end{itemize} 

        These results are particularly relevant in the context of increasing subsurface use as decarbonization efforts progress~\cite<e.g.,>{krevor2023}. Similar to oil production at Wilmington, ground deformation and fault stability (controlled by the in-situ stress state) are also critical aspects for safe and efficient carbon dioxide sequestration, hydrogen storage, and geothermal energy. Therefore, we emphasize the value of local stress measurements at different depths, formation mechanical characterization, and fault stability analysis in mitigating risk in subsurface energy technologies.

         Work is under way to extend geologically-realistic, physics-based models of reservoir operations in California to the basin scale. Such models can capture the effects of concurrent reservoir operations across multiple fields and, combined with seismicity catalogs, can be used to quantitatively assess whether stress changes from reservoir operations are associated with observed earthquakes. This effort is motivated in part by suggestions that historical earthquakes in the Los Angeles Basin may have been induced~\cite{hough2016,hough2018}, which carries direct implications for the management of subsurface energy resources in the region.

\section*{Open Research}
Seismic data were provided by the City of Long Beach, California. Other geophysical data to generate the computational mesh were obtained from sources detailed in \S~\ref{sec:stress} and \S~\ref{sec:setup-geomechanics}, and integrated into Skua-GoCad and Petrel. Hydrocarbon production and water injection data for the Wilmington Field (average monthly rates 1936-2020, latitude, and longitude per well), are available from the California Department of Conservation at \url{https://www.conservation.ca.gov}.

\acknowledgments
L.S. thanks Natasha Toghramadjian for guidance on accessing historical Wilmington documents, and Paul Segall for helpful discussions on the relationship between stress state and induced seismicity. This work was the result of a collaboration between researchers at Harvard University and MIT, largely supported by National Science Foundation (NSF) grant EAR-2141382 to Harvard University and NSF grant EAR-2141316 to MIT. This support is gratefully acknowledged. Any opinions, findings, and conclusions or recommendations expressed in this material are
those of the authors and do not necessarily reflect the views of the NSF.

\bibliography{references_master}

\end{document}